\documentclass[aps,prd,twocolumn,showpacs,preprintnumbers,nofootinbib,amsmath,amssymb]{revtex4}

\usepackage{graphicx}
\usepackage{dcolumn}
\usepackage{bm}
\usepackage{amsmath}
\usepackage{braket}
\usepackage[normalem]{ulem}
\usepackage[hidelinks]{hyperref}
\usepackage{epstopdf}
\usepackage{placeins}
\usepackage{multirow}
\usepackage{makecell}
\usepackage{cleveref}

\graphicspath{{figures/}}

\newcommand{\beq}{\begin{eqnarray}}
\newcommand{\eeq}{\end{eqnarray}}
\newcommand{\beqnn}{\begin{eqnarray*}}
\newcommand{\eeqnn}{\end{eqnarray*}}

\newcommand{\DIGA}{\mathit{DIGA}}
\newcommand{\dof}{\mathrm{dof}}

\def\spose#1{\hbox to 0pt{#1\hss}}
\def\ltapprox{\mathrel{\spose{\lower 3pt\hbox{$\mathchar"218$}}
 \raise 2.0pt\hbox{$\mathchar"13C$}}}

\begin{document}

\title{$\theta$-dependence in the small-$N$ limit of $2d$ $CP^{N-1}$ models}

\author{Mario Berni}
\email{marioberni91@gmail.com}
\affiliation{Universit\`a di Pisa and INFN Sezione di Pisa,\\ 
Largo Pontecorvo 3, I-56127 Pisa, Italy
}

\author{Claudio Bonanno}
\email{claudio.bonanno@pi.infn.it}
\affiliation{Universit\`a di Pisa and INFN Sezione di Pisa,\\ 
Largo Pontecorvo 3, I-56127 Pisa, Italy
}

\author{Massimo D'Elia}
\email{massimo.delia@unipi.it}
\affiliation{Universit\`a di Pisa and INFN Sezione di Pisa,\\ 
Largo Pontecorvo 3, I-56127 Pisa, Italy
}

\date{\today}

\begin{abstract}
We present a systematic numerical study of 
$\theta$-dependence around $\theta=0$ in the small-$N$ limit
of $2d$ $CP^{N-1}$ models, aimed at 
clarifying the possible presence of 
a divergent topological susceptibility in the 
continuum limit.
We follow a twofold strategy, based on one side on direct simulations
for $N = 2$ and $N = 3$ on lattices with correlation lengths 
up to $O(10^2)$, and on the other side on the small-$N$ extrapolation
of results obtained for $N$ up to $9$.
Based on that, we provide conclusive 
evidence for a finite topological susceptibility at 
$N = 3$, with a continuum estimate $\xi^2 \chi = 0.110(5)$.
On the other hand, results obtained for $N = 2$ are still inconclusive:
they are consistent with a logarithmically divergent 
continuum extrapolation, but do not yet exclude 
a finite continuum value, $\xi^2 \chi \sim 0.4$, with 
the divergence taking place for $N$ slightly below 2 in this case.
Finally, results obtained for the 
non-quadratic part of $\theta$-dependence, in particular for
the so-called $b_2$ coefficient, are consistent with 
a $\theta$-dependence matching that
of the Dilute Instanton Gas Approximation at the point
where $\xi^2 \chi$ diverges.
\end{abstract}

\maketitle

\section{Introduction}
\label{sec:intro}
The $CP^{N-1}$ models in two space-time dimensions have been extensively studied in the literature because they represent an interesting 
theoretical laboratory for the study of gauge theories~\cite{DAdda:1978vbw,advanced_topics_QFT,Vicari:2008jw}. As a matter of fact, many intriguing non-perturbative properties, such as confinement, the existence of field configurations with non-trivial topology and the related $\theta$-dependence are features that these models share with $4d$ Yang-Mills theories.

The Euclidean action of these models, including the topological term, can be written through a non-propagating Abelian field $A_\mu$ as
\beq\label{eq:total_action}
S(\theta)=\int d^2x\left[ \frac{N}{g}\bar{D}_\mu \bar{z}(x) D_\mu z(x) -i \theta q(x) \right],
\eeq
where $N$ is the number of components of the complex scalar field $z$, which is subject to the constraint $\bar{z}z=1$, $D_\mu=\partial_\mu+iA_\mu$ and
\beq
Q=\int d^2x \, q(x)= \frac{1}{2\pi} \epsilon_{\mu\nu} \int d^2x \, \partial_\mu A_\nu(x)
\eeq
is the topological charge. 
The $\theta$-dependent vacuum energy (density) is defined through the path integral as
\beq
E(\theta)=-\frac{\log Z(\theta)}{V}=-\frac{1}{V}\log \int [d\bar{z}][dz][dA]e^{-S(\theta)}
\eeq
and it can be parametrized in terms of the cumulants $\braket{Q^{n}}_c$ of the topological charge distribution at $\theta=0$ $P(Q)$ as:
\beq
f(\theta)\equiv E(\theta)-E(0)=\frac{1}{2}\chi\theta^2\left(1+\sum_{n=1}^{\infty} b_{2n}\theta^{2n}\right),
\label{free_energy_expansion}
\eeq
where the topological susceptibility
\beq\label{eq:def_chi}
\chi \equiv \frac{\braket{Q^2}_c}{V}\bigg\vert_{\theta=0} = \frac{\braket{Q^2}}{V}\bigg\vert_{\theta=0}
\eeq
parametrizes the leading $\theta^2$ term while the coefficients
\beq
\label{eq:def_b_2n}
b_{2n} \equiv (-1)^{n}\frac{2}{(2n+2)!}\frac{\braket{Q^{2n+2}}_c}{\braket{Q^2}}\bigg\vert_{\theta=0}
\eeq
parametrize the non-quadratic part.

One of the most interesting features of the $CP^{N-1}$ models is the possibility of performing a systematic expansion of any observable, including those 
related to $\theta$-dependence, in the inverse of the number of field components $1/N$ when $N\to\infty$ while $g$ is kept fixed~\cite{DAdda:1978vbw}; 
that closely resembles the $1/N$ expansion of $4d$ $SU(N)$ 
gauge theories for large number of colors. 
The similarities are non-trivial, since the relevant scaling quantity turns out to be $\theta/N$ in both cases~\cite{Witten:1980sp,Witten:1998uka,Rossi:2016uce,Bonati:2016tvi}, leading to the prediction that the quantities $\bar{b}_{2n}\equiv N^{2n}b_{2n}$ are finite in the large-$N$ limit:
this scaling behavior has been verified by numerical lattice simulations 
both for the $CP^{N-1}$~\cite{Bonati:2016tvi,Berni:2019bch} and for $SU(N)$ gauge theories~\cite{DelDebbio:2006yuf, Bonati:2016tvi} up to the quartic coefficient $b_2$.
From a quantitative point of view, $CP^{N-1}$ models
are more predictive, since the leading term
in the $1/N$ expansion is known~\cite{DAdda:1978vbw,DelDebbio:2006yuf,Rossi:2016uce,Bonati:2016tvi}
for 
all coefficients in the $\theta$-expansion of the free energy in Eq.~\eqref{free_energy_expansion}, and also the first subleading term in the case
of the topological susceptibility~\cite{Campostrini:1991kv}, while in the 
case of $4d$ $SU(N)$ gauge theories one has just phenomenological 
predictions, based on the spectrum of pseudoscalar mesons, 
for the leading term in the $1/N$ expansion of the
topological susceptibility~\cite{Witten:1978bc,Veneziano:1979ec}.

Numerical simulations underlined also 
some differences between the two theories. 
Indeed, while in the Yang-Mills case the large-$N$ expected scaling practically holds already for $N\ge 3$~\cite{Bonati:2016tvi}, this is not quite the case for the $CP^{N-1}$ models. Indeed, the large-$N$ limit of $b_2$ shows significant deviations from 
large-$N$ predictions even for $N \sim 50$, and an agreement with 
lattice data can be recovered only by including large 
higher-order corrections in $1/N$; a similar behavior is observed 
for the topological susceptibility~\cite{Bonanno:2018xtd,Berni:2019bch}.

Another important difference emerges when looking at the small-$N$ limit. Indeed, while this is predicted and observed
to be regular in the case of $4d$ $SU(N)$ gauge theories, a singular behavior 
is expected for $2d$ $CP^{N-1}$ approaching $N = 2$.
For instance, one expects a divergence of the topological susceptibility of the $CP^1$ theory, which can be justified in perturbation theory on the basis of the ultraviolet (UV) divergence of the instanton size distribution~\cite{Jevicki:1977yd,Forster:1977jv,Berg:1979uq,Fateev:1979dc,Richard:1981bv}
\beq\label{eq:instanton_size_distribution}
P_N(\rho)\propto\rho^{N-3}
\eeq
occurring for $N=2$. This result has been tested in many lattice studies 
and there seems to be a general consensus about the singular behavior 
of $\chi$ for $N=2$ and about its origin due to the presence 
of small instantons, see, e.g., 
Refs.~\cite{DElia:1995zja,DElia:1995wxi,Blatter:1995ik,Ahmad:2005dr,Lian:2006ky,Bietenholz:2010xg,Bietenholz:2018agd}.
However, there are still many aspects deserving a more careful 
investigation.\\

First of all, as we discuss later on, the actual verification
of the divergent behavior occurring for $N = 2$ requires to perform
a continuum limit extrapolation which is highly non-trivial,
since the divergence is expected to appear as
a logarithm of the UV cutoff, i.e., of the lattice spacing $a$, 
which could be difficult to disentangle from a regular 
power-law behavior in $a$ and requires numerical results spanning
over orders of magnitude in terms of the dimensionless correlation
length of the system.

The second issue is that 
a divergent behavior for the topological 
susceptibility has been claimed to be observed also 
for $N=3$ in Ref.~\cite{Lian:2006ky}, 
where the authors employed a standard lattice action along with 
an overlap definition of the topological charge, 
however this is in contradiction with previous 
results obtained using a different action discretization
and a geometric lattice definition of $Q$~\cite{Petcher198353}:
the origin of this discrepancy may be due to the fact that, according
to Eq.~\eqref{eq:instanton_size_distribution}, also for $N=3$ 
small instantons are expected to dominate, so that also 
in this case the continuum limit has to be handled with care.

Finally, one would like to understand whether the small-$N$ 
divergent behavior regards just the topological susceptibility or also other 
coefficients in the Taylor expansion of the free energy 
in Eq.~\eqref{free_energy_expansion}. In asymptotically free theories,
small size instantons are expected to be weakly interacting, so 
that a possible conjecture~\cite{Bonanno:2018xtd} is that 
$\theta$-dependence in the $N = 2$ limit be described 
by the Dilute Instanton Gas Approximation (DIGA):
\beq
f_{\DIGA}(\theta)=\chi\left(1-\cos\theta\right)
\eeq
with the divergence appearing just in $\chi$, while 
the $b_{2n}$ coefficients are finite and have alternate sign:
\beq
\label{eq:b2n_diga}
b_{2n}^{\DIGA}=(-1)^{n}\frac{2}{(2n+2)!} \, .
\eeq
It is worth stressing that this conjecture would lead, apart from the global divergent factor in front, to a smooth behavior in $\theta=\pi$, unlike what happens at large-$N$, where $f(\theta)$ is expected to have a cusp in this point. In principle, this is not in contrast with some recent theoretical results obtained for the $CP^{N-1}$ models using 't Hooft anomaly matching~\cite{Gaiotto:2017yup,Sulejmanpasic:2018upi}. Indeed, the anomaly matching at $\theta=\pi$ constrains $CP^{N-1}$ models with $N$ even to either spontaneously break the charge conjugation symmetry (i.e., $f(\theta)$ has a cusp in $\theta=\pi$) or to behave as a conformal field theory (i.e., the dynamically-generated mass $m\equiv1/\xi$ vanishes in $\theta=\pi$). While the former scenario is expected to be realized by theories with $N>2$, there is much evidence, both theoretical and numerical, that the $CP^1$ theory is conformal, see, e.g., Refs.~\cite{Affleck:1987ch,Affleck:1991tj,Alles:2007br,Sulejmanpasic:2020lyq}; thus no cusp in $\theta=\pi$ is expected for $N=2$. However, the study of the critical properties of the $CP^1$ model at $\theta=\pi$ points out that, in this case, $\chi$ should be divergent also in the thermodynamic limit at fixed lattice spacing, suggesting that DIGA may not be the end of the story for describing $\theta$-dependence for $N=2$, and that corrections to it may survive the continuum limit.

In any case, a near-DIGA small-$N$ behavior of $f(\theta)$ could explain why, unlike the case of $SU(N)$ gauge theories, large corrections are observed when studying the large-$N$ limit of $CP^{N-1}$ models, being the $1/N$ series at large $N$ not able the capture the peculiar small-$N$ behavior of the theory. A first evidence of near-DIGA behavior for the quartic coefficient $b_2$ for $N=2$ has been reported in Ref.~\cite{Bietenholz:2016szu}, where however no continuum extrapolation for this quantity is reported. As for higher values of $N$, no result is known for $N<9$.\\

The goal of the present work is to provide a systematic study of the small-$N$ $\theta$-dependence of $CP^{N-1}$ models by lattice simulations, 
in order to go beyond the present state of the art. 
To do so, we have attacked the problem from two different sides.
On one hand, we have performed extensive numerical simulations
for $N = 2$ and $N = 3$ considering dimensionless correlation lengths spanning over two orders of magnitude, namely reaching 
values of $\xi$ going above $10^2$, and considering various different ans{\"a}tze for the continuum extrapolation in order to fairly assess 
our systematic uncertainties on the final results.
On the other hand, we have also considered numerical simulations
for larger values of $N$, namely $N \in [4,9]$, for which
the continuum extrapolation is easier and better defined, then trying 
to obtain information on $N=2,3$ by a small-$N$ extrapolation
of these results. We have applied this double-front strategy 
to the determination of the topological susceptibility and 
of the first coefficients of the $\theta$ expansion of $f(\theta)$, 
up to $O(\theta^6)$: consistency of results obtained in the two
different ways provides a solid way to assess the reliability 
of our final statements, among which, for instance, the fact
that the topological susceptibility is finite for $N = 3$.

The paper is organized as follows. In Sec.~\ref{sec:setup} we describe the lattice setup adopted for discretizing the theory
and for determining the cumulants of the topological charge distribution, 
as well as our strategy for taking the continuum limit of our results.
In Sec.~\ref{sec:results} we present and discuss our numerical results and
finally, in Sec.~\ref{sec:final}, we draw our conclusions.

\section{Numerical setup}
\label{sec:setup}

In this section we briefly discuss various issues related to the discretization
of the model and of its observables, in particular those related to topology, 
and to the continuum extrapolation of the numerical results.

\subsection{Lattice discretization}

We discretized the action in Eq.~\eqref{eq:total_action} at $\theta=0$ on a periodic square lattice of size $L$ using the tree-level Symanzik-improved lattice discretization~\cite{Campostrini:1992ar}:
\beq\label{eq:Symanzik_improved_lattice_action_cpn}
\begin{aligned}
S_L = &-2N\beta_L\sum_{x,\mu} \left \{ c_1 \Re\left[\bar{U}_\mu(x)\bar{z}(x+\hat{\mu})z(x)\right] \right. \\
&\left. + c_2 \Re\left[\bar{U}_\mu(x+\hat{\mu})\bar{U}_\mu(x)\bar{z}(x+2\hat{\mu})z(x)\right]
 \right\},
\end{aligned}
\eeq
where $c_1=4/3$ and $c_2=-1/12$ are improvement coefficients, $\beta_L \equiv 1/g_L$ is the inverse bare coupling and $U_\mu(x)$ are the $U(1)$ gauge link variables. The adoption of the improved action cancels out logarithmic corrections to the leading $O(a^2)$ behavior of the discretization errors, where $a$ is the lattice spacing.

Being $CP^{N-1}$ models asymptotically free for all values of $N$, the continuum limit is approached as $\beta_L \to \infty$. The $a\to0$ limit can be traded for that of a diverging lattice correlation length $\xi_L$.
Our choice is for the second moment correlation length $\xi$, defined in the continuum theory in terms of the two-point correlation function of $P_{ij}(x) \equiv z_i(x) \bar{z}_j(x)$
\beq\label{eq:projector_definition}
G(x) \equiv \braket{P_{ij}(x)P_{ij}(0)}-\frac{1}{N}
\eeq
as
\beq\label{eq:def_xi}
\xi^2 \equiv \frac{1}{\int G(x)d^2x}\int G(x) \frac{\vert x\vert^2}{4} d^2 x.
\eeq
To define the lattice length $\xi_L$, we adopted the following definition~\cite{Caracciolo:1998gga}, expressed through the Fourier transform $\tilde{G}_L(p)$ of $G_L(x)$ (i.e., the discretization of Eq.~\eqref{eq:projector_definition}):
\beq\label{eq:def_xi_L}
\xi_L^2 = \frac{1}{4\sin^2\left(\pi/L\right)}\left[ \frac{\tilde{G}_L(0)}{\tilde{G}_L(2\pi/L)}-1 \right].
\eeq

\subsection{Discretization of the topological charge}

Regarding the topological charge $Q$, several equivalent lattice discretizations $Q_L$ can be adopted, all having the same continuum limit. 
However, at finite lattice spacing, these definitions and their 
correlations are related to the continuum by a finite multiplicative
renormalization $Z$~\cite{Campostrini:1988ab, Farchioni:1993jd}:
\beq
q_L \sim Z a^2 q + O(a^4) 
\eeq
where $q$ is the topological charge density.
We adopted a \emph{geometric} definition of the lattice charge~\cite{Berg:1981er, Campostrini:1992ar}, i.e., a definition with $Z=1$, meaning that it yields always an integer number for every lattice configuration. Among the several geometric definitions, we chose one which involves only the link variables~\cite{Campostrini:1992ar}:
\beq\label{eq:topo_charge_geo}
Q_U= \sum_x q_U(x) = 
\frac{1}{2\pi}\sum_{x} \Im \left\{ \log \left[\Pi_{12}(x)\right] \right\},
\eeq
where $\Pi_{\mu \nu} (x) \equiv U_\mu(x) U_\nu(x+\hat{\mu})\bar{U}_\mu(x+\hat{\nu})\bar{U}_\nu(x)$ is the plaquette operator.
Despite the fact that $Z = 1$, renormalization effects are still present,
since in general
$Q_L$ is related to the physical charge $Q$, configuration by configuration, 
by a relation like
\beq
Q=Z(\beta_L)Q_L+\eta
\eeq
where $\eta$ is a noise with zero average stemming from fluctuations 
at the UV scale. For a geometric charge, such noise appears in the form
of the so-called \emph{dislocations}~\cite{Berg:1981nw,Campostrini:1992it}, i.e., integer valued 
fluctuations at the 
scale of the UV cutoff and 
proliferating over physical contributions as the continuum
limit is approached. The presence of a non-zero $\eta$ gives rise to 
further additive renormalizations as one considers cumulants of 
the topological charge, including the topological susceptibility
(see Ref.~\cite{DElia:2003zne} for a review).

To suppress such noise, a smoothing method, such as cooling~\cite{Berg:1981nw,Iwasaki:1983bv,Itoh:1984pr,Teper:1985rb,Ilgenfritz:1985dz,Campostrini:1989dh,Alles:2000sc} 
or the gradient flow~\cite{Luscher:2009eq,Luscher:2010iy},
can be adopted. The general, underlying idea, is to perform a process of
minimization of the action, which damps at first local fluctuations at the UV scale.
It has been shown that various smoothing methods are all numerically 
equivalent~\cite{Bonati:2014tqa, Alexandrou:2015yba}
and also that the discretization chosen for the smoothing action does not need 
to coincide with the one adopted for the path-integral formulation~\cite{Alexandrou:2015yba}.
Therefore, for the sake of simplicity and numerical cheapness,
we decided to minimize the standard action 
(i.e., $c_1=1$ and $c_2=0$ in Eq.~\eqref{eq:Symanzik_improved_lattice_action_cpn})
using the cooling method, which consists of a sequence of local steps of action
minimization.

Contrary to Refs.~\cite{Bonanno:2018xtd,Berni:2019bch}, in this study we did
not adopt neither simulations 
at imaginary values of  
$\theta$ in order to improve the
signal-to-noise ratio of cumulants~\cite{Panagopoulos:2011rb,Bonati:2015sqt,Bonati:2016tvi}, nor improved algorithms in order 
to defeat the critical slowing down
of topological modes~\cite{DelDebbio:2004xh,Laio:2015era,Flynn:2015uma,Hasenbusch:2017unr}. This is due to the 
relative ease in obtaining precise determinations of the cumulants
of the topological charge for small values of $N$, even using standard
algorithms such as heat-bath or over-relaxation. 
For these reasons, the coefficients $b_{2n}$ are determined 
in this study by simply using
the definition given in Eq.~\eqref{eq:def_b_2n},
with the average taken over the path integral distribution at 
$\theta = 0$, where the choice for $Q$ is the geometrical topological charge 
in Eq.~\eqref{eq:topo_charge_geo} measured after a certain number of 
cooling steps, as discussed later on.

\subsection{Continuum limit at small $N$}
\label{sec:continuum_limit_small_N}

Since $\xi_L=\xi/a$ diverges as $1/a$ in the continuum limit, finite lattice spacing corrections can be expressed as a function of $1/\xi_L$. Since the adoption of the Symanzik-improved action in Eq.~\eqref{eq:Symanzik_improved_lattice_action_cpn} cancels out logarithmic corrections to the leading $O(a^2)$ behavior of lattice artifacts, one expects ultraviolet corrections to the lattice expectation value of a generic observable $\mathcal{O}$ to have the form
\beq \label{eq:UV_corrections}
\braket{\mathcal{O}}_{\text{\textit{latt}}}\left(\xi_L\right) = \braket{\mathcal{O}}_{\text{\textit{cont}}} + c \, \xi_L^{-2} + 
O\left(\xi_L^{-4}\right).
\eeq
However, when approaching $N\to2$ one expects, at least for topological observables, 
the presence of additional corrections coming from physical topological fluctuations of small size: such physical contributions
are neglected in the discretized theory until the lattice spacing 
is small enough, leading to an additional dependence on 
$a$, hence on $\xi_L$.

An a priori estimate of such effects can be done only with some assumptions,
nevertheless it can be a useful guide. For instance, taking
the perturbative estimate for the instanton size distribution 
reported in Eq.~\eqref{eq:instanton_size_distribution} and assuming 
that topological fluctuations are dominated by 
a non-interacting gas of small instantons and anti-instantons, 
one has that 
the number of instantons $n_I$ and anti-instantons $n_A$ are distributed as two independent Poissonians with equal mean:
\beq
\braket{n_I}=\braket{n_A}\propto \int_a^{\rho_0} P_N(\rho)d\rho = \int_a^{\rho_0} \rho^{N-3} d\rho ,
\eeq
where the integral is carried over sizes ranging from the UV scale, set by the lattice spacing $a$, up to a certain 
infrared length scale $\rho_0$, which is proportional to $La$. 
Since, with these hypotheses, 
\beq
\chi \propto \braket{ \left( n_I-n_A \right)^2 }=2 (\braket{n_I^2} - \braket{n_I}^2) = 2 \braket{n_I}
\eeq
we have the following predictions:
\beq
\label{eq:small_N_UV_corrections}
\chi \propto
\begin{cases}
\dfrac{\rho_0^{N-2}-a^{N-2}}{N-2}  &,\mbox{if } N>2; 
\\
\\
\log\left(\dfrac{\rho_0}{a}\right) &,\mbox{if } N=2.
\end{cases}
\eeq
Ultraviolet corrections predicted by Eqs.~\eqref{eq:small_N_UV_corrections} become negligible as $N$ grows, in particular one expects them to disappear for $N\ge4$, where the contribution from small instantons becomes negligible. 
On the other hand, the contribution of small instantons becomes dominant for $N=2$ and $3$, where it leads either to a logarithmically divergent continuum limit for $N = 2$, or to linear, instead of quadratic, corrections in the lattice spacing for $N = 3$.
Notice that, under the assumptions of independent Poisson distributions
for $n_I$ and $n_A$, the $b_{2n}$ coefficients are finite with values
as predicted by DIGA in Eq.~\eqref{eq:b2n_diga}, so that no further
corrections are expected, with respect to 
Eq.~\eqref{eq:UV_corrections}, in their approach to the continuum limit.

Such considerations will be used with caution in the following.
In particular, in order to correctly assess our systematics on the 
continuum limit, we will consider the possible presence of 
generic power law corrections in the lattice spacing 
for $N \leq 4$, both for $\chi$ and for the other terms in the 
$\theta$-expansion.

\subsection{Continuum limit and smoothing}
\label{sec:continuum_smoothing}

Since we adopt a smoothing method in order to remove field fluctuations
at the UV cutoff scale, which are responsible for unphysical contributions
to lattice topological observables, we need to fix how the amount
of smoothing is changed as one approaches the continuum limit.

Smoothing algorithms work in general as diffusive processes, affecting 
field correlations up to a given distance, i.e., up to 
a given {\em smoothing radius}, which is proportional, in 
dimensionless lattice units, to the square root of the amount of smoothing, 
i.e., to $\sqrt{n_{cool}}$ for cooling, where $n_{cool}$ is the number of 
cooling steps, or to $\sqrt{t}$ for the gradient flow, where $t$ is 
the flow time. It is a standard procedure to change the amount of smoothing
so that the smoothing radius is kept fixed in physics units: that 
would correspond to change $n_{cool}$ proportionally to $\xi_L^2$.
However, for a model like the one we are exploring, where small
distance physical contributions are expected to be quite significant, 
one should be careful and take care,
additionally, of the dependence of continuum results on the physical smoothing 
radius, eventually sending it to zero.

An alternative to this difficult 
double limit procedure is to take the continuum
limit at fixed number of cooling steps $n_{cool}$, 
so that the smoothing radius goes to zero proportionally to $a$ and there
is no possibility that physical contributions at small scales are smoothed 
away. 
There are 
good reasons to believe that such procedure works 
correctly in the present case. 

As we have discussed above, for a geometric
charge like the one used in this study renormalization effects are 
essentially due to dislocations leading to a wrong and/or ambiguous 
counting of the topological winding number. Such dislocations consist 
of exceptional field fluctuations
living at the lattice spacing scale, hence it is reasonable to 
expect that they will be suppressed by a given and fixed number of 
cooling steps $n_{cool}$, independently of the value of the 
lattice spacing $a$.

Based on such considerations, in the following we will consider 
results obtained by performing the continuum limit at fixed $n_{cool}$,
then carefully checking the possible systematics related to this procedure. 
In particular, 
we will show that while results obtained at finite lattice spacing,
but at different number of cooling steps, usually differ from 
each other well beyond their statistical errors,
the so obtained continuum-extrapolated results
are not significantly dependent on $n_{cool}$, and usually well
within statistical errors.

\section{Numerical results}
\label{sec:results}

In~\Cref{tab:info_N_2,tab:info_N_3,tab:info_N_4,tab:info_N_5,tab:info_N_6,tab:info_N_7,tab:info_N_8} we summarize the parameters of the performed simulations, along with the total accumulated statistics. Configurations were generated using standard local algorithms, in particular our elementary Monte Carlo step consisted in 4 lattice sweeps of over-relaxation followed by a sweep of over-heath-bath; measures were taken every 10 Monte Carlo steps.\\
We simulated $CP^{N-1}$ models with $N$ ranging from $2$ to $8$. For each value of $N$ we simulated several runs at different values of the correlation length (i.e., at different values of $\beta_L$); for each $\xi_L$ we measured $\xi^2 \chi$, $b_2$ and $b_4$ in order to be able to extrapolate these quantities towards the continuum limit.

\begin{figure}[!htb]
\centering
\includegraphics[scale=0.52]{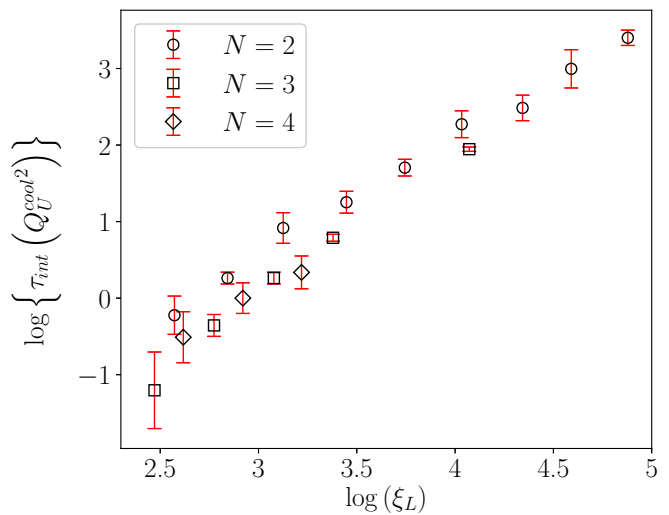}
\caption{Behavior of the integrated auto-correlation time $\tau_{int}$ of ${Q_U^{cool}}^2$, expressed in units of the Monte Carlo updating step defined in the text, as a function of $\xi_L$ and $N$ for $\xi_L>10$. The integrated auto-correlation time was computed through a standard binned bootstrap analysis on samples obtained after $n_{cool}=50$ cooling steps.}
\label{fig:tau}
\end{figure}

\begin{figure}[!htb]
\centering
\includegraphics*[scale=0.52]{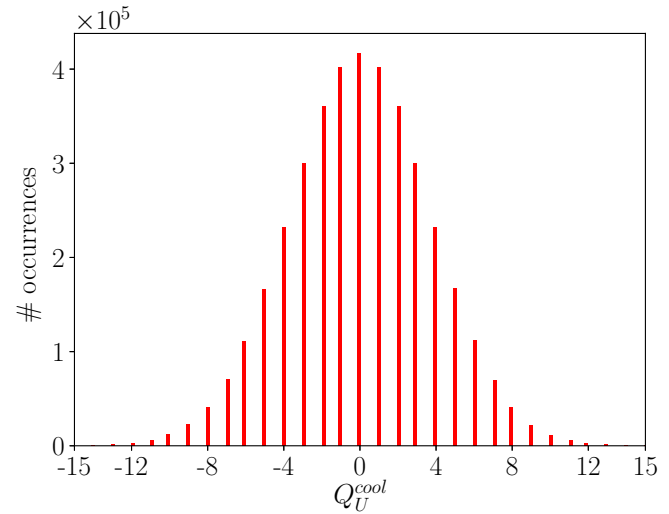}	
\caption{Distribution of the lattice topological charge $Q_U^{cool}$ during a run with $N=2$, $\beta=1.65$ and $L=1024$ and measured after $n_{cool}=50$ cooling steps.}
\label{fig:MC_evolution_Q}
\end{figure}

As already anticipated, topological freezing is not an issue at small $N$: as a matter of fact, standard local updates allowed a reliable sampling of the topological charge in all simulations, even for the largest explored values of the correlations length. In Fig.~\ref{fig:tau} we show, as an example, the integrated auto-correlation time $\tau_{int}$ of ${Q_U^{cool}}^2$, 
in units of the Monte Carlo step defined above, as a function of $\xi_L$ for $\xi_L>10$ and $N=2,3$ and $4$; $\tau_{int}$ was obtained by a binned bootstrap using the standard procedure described, e.g., in Ref.~\cite{DelDebbio:2004xh}. As it can be appreciated, $\tau_{int}$ is in all cases much smaller than the total collected statistics; indeed, in all our simulations we observed many fluctuations of $Q_U^{cool}$ during the Monte Carlo evolution. It is interesting to notice that, contrary to what happens at large $N$~\cite{DelDebbio:2004xh,Bonanno:2018xtd}, $\tau_{int}$ diverges
as a power law in $\xi_L$ (rather than exponentially), so that the critical
slowing down of the topological charge is a much milder problem in this case: this is likely
due to the presence of small instantons, which are easier to decorrelate. 
In Fig.~\ref{fig:MC_evolution_Q} we show, as an example, the distribution 
of $Q_U^{cool}$ obtained for $N=2$ and at the largest explored values of $\xi_L$ and $L$.

\begin{figure}[!htb]
\hspace*{-0.4cm}
\includegraphics[scale=0.52]{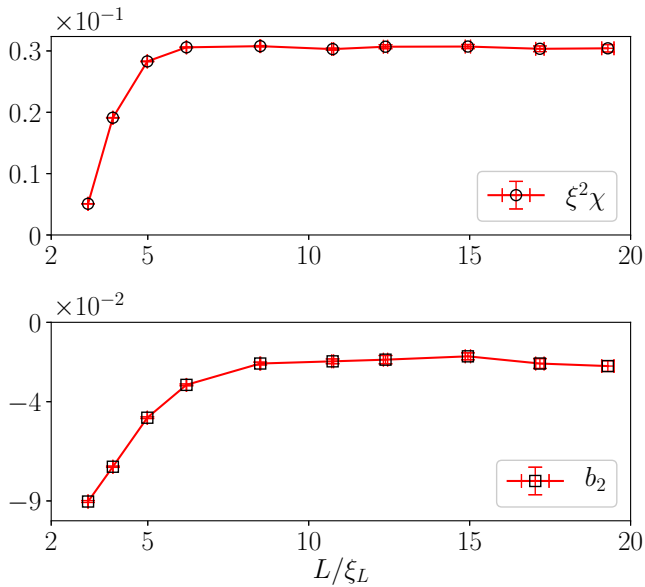}	
\caption{Example of finite size scaling of $\xi^2 \chi$ and $b_2$ as a function of $L/\xi_L$ for $N=6$ and $\beta_L=0.80$ with measures taken after $n_{cool}=50$ cooling steps.}
\label{fig:finite_size_effects}
\end{figure}

\begin{figure}[!htb]
\hspace*{-0.55cm}
\includegraphics[scale=0.54]{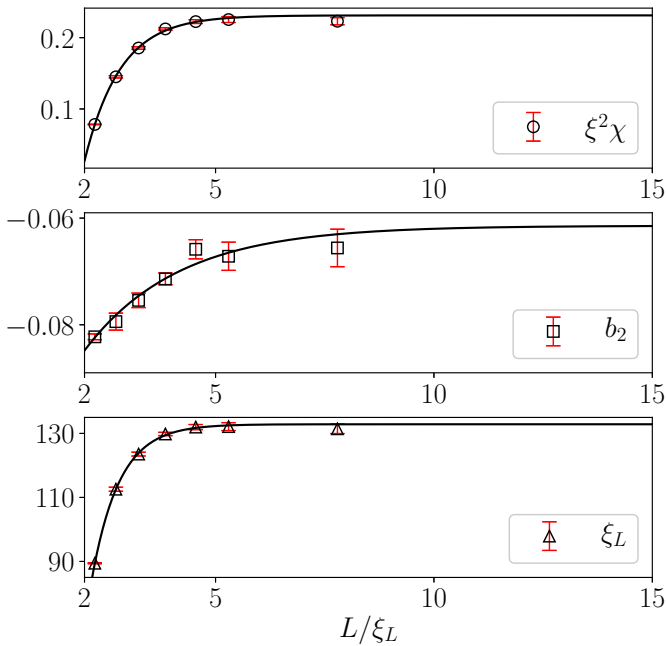}	
\caption{Examples of extrapolation towards the infinite volume limit of $\xi^2\chi$, $b_2$ and $\xi_L$ for $N=2$ and $\beta_L=1.65$ with measures taken after $n_{cool}=50$ cooling steps. Best fits of Eq.~\eqref{eq:finite_size_scaling} performed with 4 degrees of freedom (dof) yield, respectively, $\chi^2/\dof=1.4, 0.99, 0.6$.}
\label{fig:thermo_limit}
\end{figure}

Concerning the choice of the lattice size, we performed simulations at fixed $L/\xi_L$, ensuring that $L/\xi_L \sim 12$-$15$ for each run to have finite size effects under control~\cite{Rossi:1993nz}, see Fig.~\ref{fig:finite_size_effects}. In those cases when that was not possible (more precisely, for high-$\xi_L$ runs for $N=2$), several lattices with different sizes were simulated and then the infinite volume limit was performed by fitting the $L$-dependence of every observable $\mathcal{O}$ to the law
\beq\label{eq:finite_size_scaling}
\mathcal{O}(L) = \mathcal{O}_\infty \left(1-a\,e^{-bL/\xi_L}\right),
\eeq
where $\mathcal{O}_\infty$ is the desired quantity and $a$ and $b$ are additional fit parameters. 
An example of extrapolation towards the thermodynamic limit is shown in Fig.~\ref{fig:thermo_limit}.
\begin{table}[!htb]
\centering
\begin{center}
\begin{tabular}{|c|c|c|c|c|}
\hline
&&&&\\[-1em]
$\beta_L$ & $L$ & $\xi_L$ & $L/\xi_L$ & Stat. (M) \\
\hline
&&&&\\[-1em]
0.80 & 22   & 1.704(13)  & 12.9 & 3.2  \\
0.85 & 26   & 2.037(16)  & 12.8 & 3.2  \\
0.90 & 30   & 2.4764(72) & 12.1 & 20.8 \\
0.95 & 36   & 3.0309(28) & 12.0 & 201  \\
1.00 & 46   & 3.7726(48) & 12.3 & 130  \\
1.05 & 58   & 4.7345(59) & 12.3 & 150  \\
1.10 & 74   & 6.0142(91) & 12.2 & 114  \\
1.15 & 94   & 7.735(16)  & 12.3 & 66.7 \\
1.20 & 122  & 9.958(51)  & 12.4 & 12.2 \\
1.25 & 160  & 13.10(14)  & 12.2 & 3.4  \\
1.30 & 210  & 17.16(19)  & 12.2 & 3.9  \\
1.35 & 280  & 22.77(34)  & 12.3 & 3.6  \\
\hline
&&&&\\[-1em]
\multirow{4}{*}{1.40} & 98   & 29.031(40) & 3.4  & 3.5 \\
                      & 198  & 31.115(55) & 6.4  & 13.5 \\
                      & 298  & 31.32(19)  & 9.5  & 6.0 \\
                      & 396  & 31.41(50)  & 12.6 & 3.4 \\
\hline
&&&&\\[-1em]
\multirow{4}{*}{1.45} & 122  & 38.167(55) & 3.2  & 3.5 \\
                      & 240  & 41.352(89) & 5.8  & 8.7 \\
                      & 360  & 41.48(28)  & 8.7  & 4.0 \\
                      & 480  & 42.30(76)  & 11.3 & 2.2 \\
\hline
&&&&\\[-1em]
\multirow{4}{*}{1.50} & 168  & 51.43(11)  & 3.3  & 1.5 \\
                      & 336  & 55.75(21)  & 6.0  & 4.5 \\
                      & 504  & 57.26(78)  & 8.8  & 1.9 \\
                      & 672  & 56.4(1.5)  & 11.9 & 2.0 \\
\hline
&&&&\\[-1em]
\multirow{5}{*}{1.55} & 214  & 67.95(21)  & 3.1  & 1.5 \\
                      & 400  & 74.97(31)  & 5.3  & 3.8 \\
                      & 500  & 74.60(57)  & 6.7  & 2.4 \\
                      & 600  & 78.0(1.0)  & 7.7  & 1.6 \\
                      & 700  & 76.9(1.7)  & 9.1  & 1.2 \\
\hline
&&&&\\[-1em]
\multirow{5}{*}{1.60} & 260  & 88.13(34)  & 3.0  & 1.1 \\
                      & 400  & 97.92(33)  & 4.1  & 3.8 \\
                      & 500  & 98.95(59)  & 5.1  & 2.4 \\
                      & 600  & 102.23(94) & 5.9  & 1.6 \\
                      & 700  & 98.6(1.5)  & 7.1  & 1.2 \\
\hline
&&&&\\[-1em]
\multirow{7}{*}{1.65} & 200  & 89.44(16)  & 2.2  & 2.4 \\
                      & 306  & 112.57(63) & 2.7  & 0.77 \\
                      & 400  & 123.54(60) & 3.2  & 1.5 \\
                      & 500  & 129.82(52) & 3.9  & 3.7 \\
                      & 600  & 131.98(80) & 4.5  & 2.5 \\
                      & 700  & 132.1(1.2) & 5.3  & 1.8 \\
                      & 1024 & 131.4(1.5) & 7.8  & 6.3 \\
\hline
\end{tabular}
\end{center}
\caption{Simulations summary for $N=2$. Statistics is expressed in millions (M) and every measure is taken after 10 elementary Monte Carlo steps (see the text for further details).}
\label{tab:info_N_2}
\end{table}
\begin{table}[!htb]
\centering
\begin{center}
\begin{tabular}{|c|c|c|c|c|}
\hline
&&&&\\[-1em]
$\beta_L$ & $L$ & $\xi_L$ & $L/\xi_L$ & Stat. (M) \\
\hline
&&&&\\[-1em]
0.70  & 18  & 1.4797(59) & 12.2 & 3.5  \\
0.75  & 22  & 1.8227(76) & 12.1 & 3.5  \\
0.80  & 28  & 2.298(10)  & 12.2 & 3.5  \\
0.85  & 36  & 2.906(14)  & 12.4 & 3.5  \\
0.90  & 46  & 3.750(31)  & 12.3 & 1.5  \\
0.95  & 58  & 4.930(23)  & 11.8 & 3.5  \\
0.975 & 68  & 5.710(29)  & 11.9 & 3.2  \\
1.00  & 80  & 6.588(25)  & 12.1 & 7.8  \\
1.025 & 92  & 7.589(28)  & 12.1 & 8.3  \\
1.05  & 106 & 8.828(32)  & 12.1 & 8.3  \\
1.075 & 122 & 10.231(37) & 11.9 & 8.8  \\
1.10  & 146 & 11.834(39) & 12.3 & 13.5 \\
1.15  & 200 & 16.019(55) & 12.5 & 15.5 \\
1.20  & 264 & 21.717(71) & 12.2 & 19.4 \\
1.25  & 374 & 29.342(82) & 12.7 & 37.8 \\
1.366 & 720 & 58.69(19)  & 12.3 & 44.9 \\
\hline
\end{tabular}
\end{center}
\caption{Simulations summary for $N=3$.}
\label{tab:info_N_3}
\end{table}
\begin{table}[!htb]
\centering
\begin{center}
\begin{tabular}{|c|c|c|c|c|}
\hline
&&&&\\[-1em]
$\beta_L$ & $L$ & $\xi_L$ & $L/\xi_L$ & Stat. (M) \\
\hline
&&&&\\[-1em]
0.70 & 22  & 1.7842(59) & 12.3 & 2.8 \\
0.75 & 30  & 2.3137(97) & 13.0 & 2.8 \\
0.80 & 38  & 3.039(12)  & 12.5 & 3.1 \\
0.85 & 50  & 4.009(17)  & 12.5 & 3.1 \\
0.90 & 66  & 5.398(22)  & 12.2 & 3.1 \\
0.95 & 90  & 7.314(33)  & 12.3 & 3.1 \\
1.00 & 120 & 9.997(45)  & 12.0 & 3.1 \\
1.05 & 170 & 13.700(71) & 12.4 & 3.1 \\
1.10 & 226 & 18.55(10)  & 12.2 & 3.5 \\
1.15 & 320 & 25.00(20)  & 12.8 & 3.5 \\
\hline
\end{tabular}
\end{center}
\caption{Simulations summary for $N=4$.}
\label{tab:info_N_4}
\end{table}
\begin{table}[!htb]
\centering
\begin{center}
\begin{tabular}{|c|c|c|c|c|}
\hline
&&&&\\[-1em]
$\beta_L$ & $L$ & $\xi_L$ & $L/\xi_L$ & Stat. (M) \\
\hline
&&&&\\[-1em]
0.70 & 32  & 2.1263(98) & 15.0 & 3.5 \\
0.75 & 44  & 2.830(13)  & 15.5 & 4.6 \\
0.80 & 58  & 3.882(20)  & 14.9 & 3.5 \\
0.85 & 84  & 5.270(35)  & 15.9 & 3.5 \\
0.90 & 106 & 7.149(41)  & 14.8 & 3.5 \\
0.95 & 146 & 9.849(59)  & 14.8 & 3.5 \\
1.00 & 198 & 13.40(11)  & 14.8 & 2.2 \\
1.05 & 278 & 18.117(93) & 15.3 & 7.4 \\
1.10 & 368 & 24.77(19)  & 14.9 & 3.6 \\
\hline
\end{tabular}
\end{center}
\caption{Simulations summary for $N=5$.}
\label{tab:info_N_5}
\end{table}
\begin{table}[!htb]
\centering
\begin{center}
\begin{tabular}{|c|c|c|c|c|}
\hline
&&&&\\[-1em]
$\beta_L$ & $L$ & $\xi_L$ & $L/\xi_L$ & Stat. (M) \\
\hline
&&&&\\[-1em]
0.60 & 18  & 1.4035(36) & 12.8 & 2.6 \\
0.65 & 24  & 1.8608(53) & 12.9 & 2.6 \\
0.70 & 32  & 2.5032(76) & 12.8 & 2.5 \\
0.75 & 42  & 3.394(11)  & 12.4 & 2.2 \\
0.80 & 58  & 4.671(15)  & 12.4 & 2.4 \\
0.85 & 78  & 6.389(21)  & 12.2 & 2.6 \\
0.90 & 108 & 8.760(31)  & 12.3 & 2.6 \\
0.95 & 150 & 11.955(48) & 12.5 & 2.8 \\
1.00 & 200 & 16.112(41) & 12.4 & 6.5 \\
\hline
\end{tabular}
\end{center}
\caption{Simulations summary for $N=6$.}
\label{tab:info_N_6}
\end{table}
\begin{table}[!htb]
\centering
\begin{center}
\begin{tabular}{|c|c|c|c|c|}
\hline
&&&&\\[-1em]
$\beta_L$ & $L$ & $\xi_L$ & $L/\xi_L$ & Stat. (M) \\
\hline
&&&&\\[-1em]
0.70 & 44  & 2.876(12)  & 15.3 & 2.2 \\
0.75 & 58  & 3.940(19)  & 14.7 & 1.7 \\
0.80 & 80  & 5.403(26)  & 14.8 & 2.4 \\
0.85 & 112 & 7.436(35)  & 15.1 & 3.1 \\
0.90 & 154 & 10.096(48) & 15.3 & 3.4 \\
0.95 & 210 & 13.717(68) & 15.3 & 3.6 \\
1.00 & 304 & 18.76(11)  & 16.2 & 3.9 \\
\hline
\end{tabular}
\end{center}
\caption{Simulations summary for $N=7$.}
\label{tab:info_N_7}
\end{table}
\begin{table} [!htb]
\centering
\begin{center}
\begin{tabular}{|c|c|c|c|c|}
\hline
&&&&\\[-1em]
$\beta_L$ & $L$ & $\xi_L$ & $L/\xi_L$ & Stat. (M) \\
\hline
&&&&\\[-1em]
0.65 & 36  & 2.3175(76) & 15.5 & 3.0 \\
0.70 & 48  & 3.232(12)  & 14.8 & 2.2 \\
0.75 & 68  & 4.452(17)  & 15.3 & 3.0 \\
0.80 & 92  & 6.056(27)  & 15.2 & 2.4 \\
0.85 & 124 & 8.242(42)  & 15.0 & 2.1 \\
0.90 & 170 & 11.306(52) & 15.0 & 2.7 \\
0.95 & 240 & 15.164(73) & 15.8 & 3.4 \\
\hline
\end{tabular}
\end{center}
\caption{Simulations summary for $N=8$.}
\label{tab:info_N_8}
\end{table}

\subsection{Results for $\xi^2 \chi$, $N > 3$}
\label{subsection_results_continuum_limit_chi}

\begin{figure}[htb!]
\hspace*{-0.7cm}
\includegraphics[scale=0.52]{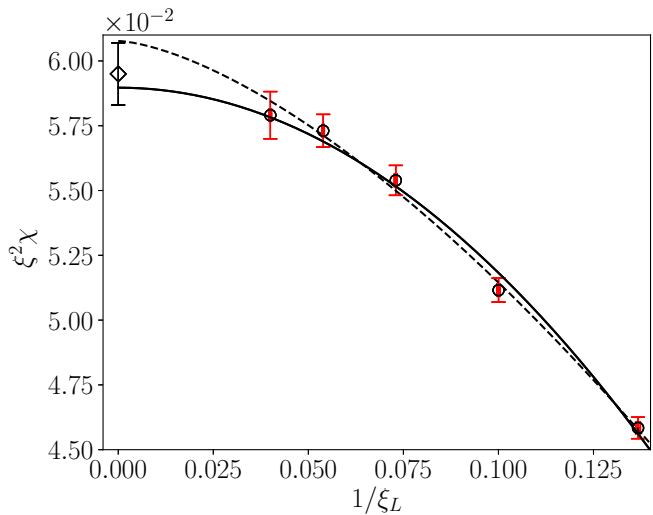}
\caption{Extrapolation towards the continuum limit of $\xi^2 \chi$ for $N=4$ using data in the range $\xi_L>6$ taken after $n_{cool}=50$ cooling steps. The solid line represents the best fit obtained using fit function $f(x)=a_0+a_1x^2$ (where $x=1/\xi_L$) while the dashed line represents the one obtained with $f(x)=a_0+a_3x^c$. In the latter case the exponent is $c=1.52(37)$ and best fits yield, respectively, $\chi^2/\dof=2.97/3$ and $\chi^2/\dof=1.34/2$. The diamond point represents our final estimation, see the text for more details on the assessment of systematic errors.}
\label{fig:continuum_limit_N_4_chi}
\end{figure}

First, we consider the case $N>3$, for which a finite continuum limit is surely expected for the topological susceptibility, with no qualitative differences in the continuum scaling compared to the large-$N$ case. For this reason,
in order to extrapolate the quantity $\xi^2\chi$ towards the continuum limit 
we have fitted its dependence on $\xi_L$ according to the ansatz
\beq\label{eq:fit_function_N>3}
f(x) = a_0 + a_1 x^2 + a_2 x^4, \quad x=1/\xi_L.
\eeq
Only for $N = 4$, due to its proximity to $N = 2$ and $3$ (see later discussion),
we have also considered the possible presence of further 
power law corrections. 
An example of continuum extrapolation is depicted, for $N=4$, in Fig.~\ref{fig:continuum_limit_N_4_chi}.

Several sources of systematic errors have been checked: first, the extrapolation was performed fitting data in several ranges of $\xi_L$ to check that the obtained extrapolations were all consistent with each other and that, as the fit range is restricted, the $O(x^4)$ corrections became negligible. Second, 
as anticipated in Sec.~\ref{sec:continuum_smoothing}, 
we extrapolated the continuum limit at fixed number of cooling step $n_{cool}$ for several values of $n_{cool}$, checking 
that this procedure does not introduce significant systematics 
in continuum-extrapolated values.

\begin{figure}[htb!]
\hspace*{-0.4cm}
\includegraphics[scale=0.52]{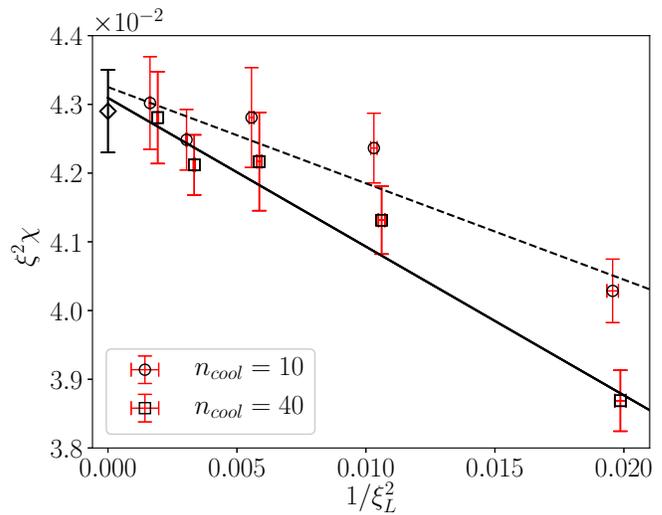}
\caption{Extrapolation towards the continuum limit of $\xi^2 \chi$ for $N=5$ using data in the range $\xi_L>7$. The solid line represents the best fit obtained using fit function $f(x)=a_0+a_1x^2$ (where $x=1/\xi_L$) with measures taken after $n_{cool}=40$ cooling steps while the dashed line represents the one obtained with the same fit function with $n_{cool}=10$ cooling steps. The best fits yield, respectively, $\chi^2/\dof=1.68/3$ and $\chi^2/\dof=2.25/3$. Data obtained for different numbers of cooling steps have been plotted slightly shifted to improve readability. The diamond point represents our final estimation of the continuum limit.}
\label{fig:cooling_systematics_continuum_limit_N_5_chi}
\end{figure}

\begin{figure}[htb!]
\hspace*{-0.5cm}
\includegraphics[scale=0.52]{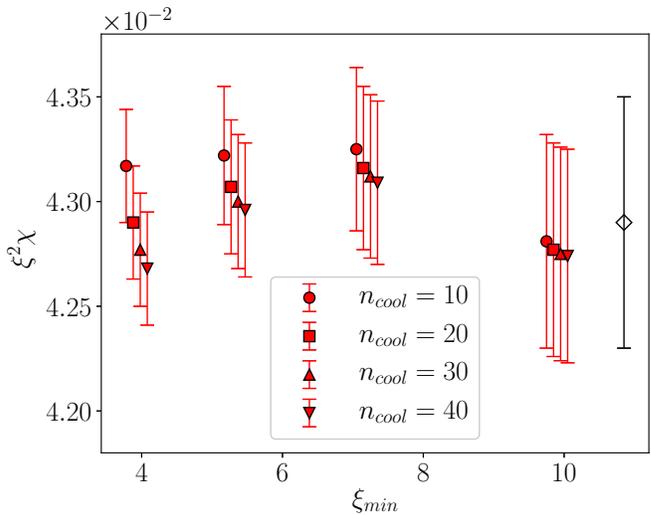}
\caption{Example of study of systematic errors on the continuum extrapolation of $\xi^2 \chi$ for $N=5$ and for 4 different number of cooling steps $n_{cool}$. Extrapolations obtained for different values of $n_{cool}$ are plotted slightly shifted to improve readability. This extrapolations are all obtained using the fit function $f(x)=a_0+a_1x^2$, where $x=1/\xi_L$. The diamond point represents our final estimation of the continuum limit.}
\label{fig:systematics_continuum_limit_N_5_chi}
\end{figure}

When changing the number of cooling steps, we observe that, while measures at coarse lattice spacing differ, the dependence on $n_{cool}$ is less and less visible as the continuum limit is approached, making the continuum extrapolation stable as $n_{cool}$ is varied. In Fig.~\ref{fig:cooling_systematics_continuum_limit_N_5_chi} we show an example of the continuum extrapolation at two different values of $n_{cool}$ for $N=5$ while, in Fig.~\ref{fig:systematics_continuum_limit_N_5_chi}, we show, for the same case, that any variation in the continuum extrapolation observed when changing $n_{cool}$ is well contained inside our statistical errors.
The final, continuum determinations obtained for $\xi^2 \chi$ are 
reported in Tab.~\ref{tab:chi_vs_N}. 

Some of our results can be compared
with previous literature. For instance, the case $N = 4$ was studied also 
in Ref.~\cite{Campostrini:1992it}, reporting a continuum extrapolation 
$\xi^2 \chi(N=4)=0.06$ with an error of the order of 10 \%, which is in agreement 
with our present result $\xi^2\chi(N=4)=0.0595(12)$,
even if with a larger uncertainty. For $N = 6$ one can find 
a previous determination in Ref.~\cite{Lian:2006ky}: even if no
continuum extrapolation is reported there, the result at the 
smallest explored lattice spacing,
$\xi^2\chi(N=6)\simeq 0.037(6)$, is consistent with our continuum extrapolation,
$\xi^2\chi(N=6)=0.0338(3)$. In both cases, the increased accuracy of our determinations 
is mostly due to the larger statistics and/or number of simulation points adopted for 
the continuum extrapolation.

\subsection{Results for $\xi^2 \chi$, $N = 3$}
\label{sec:N_3_results}

In the $N=3$ case
we fit the dependence of $\xi^2 \chi$ on $\xi_L$
according to the following function:
\beq\label{eq:fit_function_N=3}
g(x)= a_0+a_1x^2+a_2x^4+a_3x^c, \quad x=1/\xi_L,
\eeq
where we consider both the extrapolations obtained 
with fixed $c=1$, as suggested by the 
ansatz in Eq.~\eqref{eq:small_N_UV_corrections},
and with $c$ left as a free parameter.
In both cases, the ansatz in Eq.~\eqref{eq:fit_function_N=3} 
provides a very good description of our numerical results, 
and the $x^4$ corrections turn out to be necessary
only when considering in the fit range correlation 
lengths as small as $\xi_L \lesssim 5$.
Moreover, even when $c$ is treated as a free parameter, its values
turn out to be compatible with 1 within errors, thus giving
further numerical support to the ansatz 
in Eq.~\eqref{eq:small_N_UV_corrections}.

\begin{figure}[htb!]
\hspace*{-0.4cm}
\includegraphics[scale=0.52]{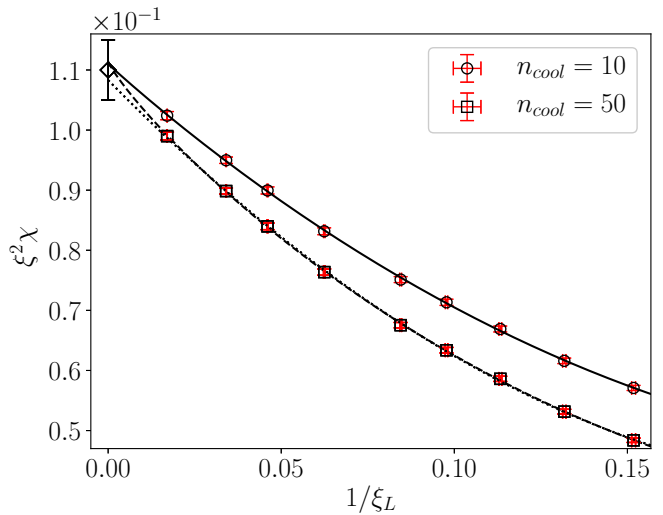}
\caption{Extrapolation towards the continuum limit of $\xi^2 \chi$ for $N=3$ using data in the range $\xi_L>6$. The dashed and dotted lines represent, respectively, the best fits obtained for $n_{cool} = 50$ using fit function $g(x)=a_0+a_1x^2+a_3x^c$ (where $x=1/\xi_L$) either leaving $c$ as a free parameter ($\chi^2/\dof=2.4/6$) or with fixed $c = 1$ ($\chi^2/\dof=1.3/5$); in the latter case the free exponent turns $c=0.85(14)$.
The solid line, instead, represents the best fit obtained with $c$
as a free parameter for
$n_{cool}=10$ cooling steps: in this case $c=0.98(18)$ 
($\chi^2/\dof=1.35/5$). The diamond point represents our final estimation of the continuum limit, which takes into account all systematics related to
the variability in fitting function, fitting range and number 
of cooling steps.} 
\label{fig:continuum_limit_N_3_chi}
\end{figure}

Examples of continuum extrapolations 
are shown in Fig.~\ref{fig:continuum_limit_N_3_chi}, where
we also report our final continuum estimate for $N = 3$,
$\xi^2 \chi = 0.110(5)$. The quoted error includes all possible systematics
related to the variability of the fit parameter $a$ 
when changing the fitting ansatz, 
the fitting range (with $\xi_{min}$ varied between 2 and 7)
and the number of cooling 
steps (with $n_{cool}$ varied between 10 and 50). 

Also in this case, a comparison with previous literature is appropriate and interesting.
Early results obtained in Ref.~\cite{Petcher198353}, 
$\xi^2 \chi(N=3)\simeq 0.09$, were not far from our present estimate,
even if no error was quoted in that case. However, later results
pointed out to a possible wrong scaling of $\chi$ towards the continuum limit,
hence to a possible divergence of $\chi$ even for $N = 3$~\cite{Lian:2006ky}.
Our present results show that $\xi^2 \chi$ is in fact finite for $N = 3$, 
even if the continuum limit extrapolation needs special care, because 
of the small instanton contributions. Since this point was debated in 
previous literature, it is important to stress that our continuum
extrapolation for $N = 3$ is fully confirmed by the small-$N$ extrapolation 
based on $N > 3$ results that we present in section~\ref{sec:small_N_limit_chi}.

Let us now discuss more in detail the systematics related to cooling.
Fig.~\ref{fig:continuum_limit_N_3_chi} shows that results obtained 
for $n_{cool} = 10$ and $50$ are significantly different from each other,
nevertheless, 
as also evident for one particular fit ansatz in 
Fig.~\ref{fig:continuum_limit_N_3_chi}, their continuum limit shows
very little variations, when compared to statistical errors 
on the fit parameters.
There is a simple way to understand why results at different 
$n_{cool}$ differ so much at finite $\xi_L$, while coinciding 
in the $\xi_L \to \infty$ limit. 
As already discussed in Sec.~\ref{sec:continuum_smoothing},
cooling acts as a diffusive process which smooths away field fluctuations
(both physical and unphysical) below an effective radius
$r = a \hat r (n_{cool})$, where the radius in lattice spacing units,
$\hat r$, is a function
of $n_{cool}$ only, i.e.~independent of the lattice spacing $a$. 
At fixed lattice spacing, different values of $n_{cool}$
lead to different values of $r$ hence to different values of the topological
susceptibility, because a different amount of physical signal below 
$r$ is removed: the effect can be particularly significant for 
small $N$, where topological fluctuations at small scales are more
abundant.
On the other hand, the effect must fade away as $a \to 0$.

\begin{figure}[htb!]
\hspace*{-0.45cm}
\includegraphics[scale=0.52]{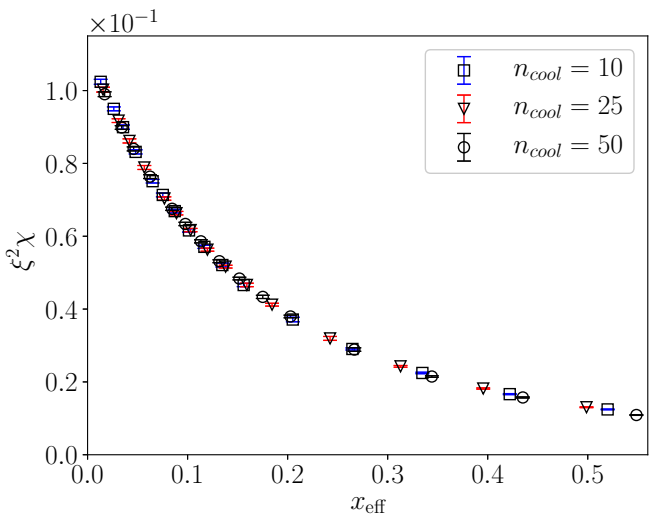}
\caption{
Results obtained for $\xi^2 \chi$ in the $N = 3$ model and for
different number of cooling steps, plotted as a function of an effective 
variable $x_{\rm eff}$ proportional to 
$\hat r(n_{cool})/ \xi_L = r / \xi$.
The perfect collapse of data proves that the dependence of 
$\xi^2 \chi$ on $n_{cool}$ can be interpreted in terms of varying 
effective radius $r = a \hat r(n_{cool})$ below which 
the topological signal is lost, and as such disappears after 
continuum extrapolation $a \to 0$. The exact value of 
$x_{\rm eff}$ has been fixed conventionally to 
$x_{\rm eff} = (\hat r (n_{cool})/\xi_L) / \hat r(n_{cool} = 50)$ and we 
have found empirically $\hat r(25) / \hat r(50) \simeq 1.1$ and 
$\hat r(10) / \hat r(50) \simeq 1.3$.
\label{fig:cool_scaling}}
\end{figure}

If the above picture is correct, results obtained
at different $a$ and different $n_{cool}$, but such that 
$r = a \hat r (n_{cool})$ is the same, should coincide:
for instance, results shown in Fig.~\ref{fig:continuum_limit_N_3_chi}
for $n_{cool} = 10$ should go onto those at $n_{cool} = 50$ 
if they are shifted along 
the horizontal axis by a constant factor 
equal to $\hat r(10) / \hat r(50)$. Such experiment has been performed
for $n_{cool} = 10, 25, 50$
in Fig.~\ref{fig:cool_scaling}, showing that it works perfectly:
plotting all data as a function of an effective 
variable $x_{\rm eff}$ proportional to $r/\xi = \hat r (n_{cool}) / \xi_L$,
where $\hat r(n_{cool})$ has been found empirically,
data at different values of $n_{cool}$ collapse perfectly onto each other
over the whole range of explored correlation lengths;
a similar collapse can be obtained also for other values of $N$.
Therefore, the dependence of $\xi^2 \chi$ on $n_{cool}$ observed  
at finite lattice spacing can indeed be simply interpreted in terms
of a global effective rescaling of $1/\xi_L$, so that such dependence 
naturally fades away when $\xi_L \to \infty$.

To summarize, our results for $N = 3$ provide solid evidence 
that $\xi^2 \chi$ is indeed finite in the continuum limit 
of this theory. On the other hand, 
that will be further checked and supported by 
an independent small-$N$ extrapolation based on $N>3$
results only, which is 
discussed in Sec.~\ref{sec:small_N_limit_chi}
and will make the evidence conclusive.

\subsection{Results for $\xi^2 \chi$, $N = 2$}
\label{sec:N_2_results}

For the $N=2$ case two possibilities 
are open. The ansatz based 
on Eq.~\eqref{eq:small_N_UV_corrections} could be correct, 
leading to a susceptibility which diverges logarithmically
in the continuum limit, i.e.~has the following dependence 
on $x = 1/\xi_L$:
\beq
\xi^2 \chi (\xi_L) = a_0^\prime \log(x / a_1^\prime) + a_2^\prime x^2 + a_3^\prime x^4. 
\label{eq:n2divergent}
\eeq
On the other hand, corrections to such prediction
leading to a finite $\xi^2 \chi$ cannot be excluded a priori,
in this case one should consider a dependence 
on $\xi_L$ like that used in 
Eq.~\eqref{eq:fit_function_N=3} for the $N=3$ case,
i.e.:
\beq
\xi^2 \chi (\xi_L) = a_0 + a_1 x^2 + a_2 x^4 + a_3 x^c. 
\label{eq:n2finite}
\eeq
where $c$ is a positive exponent.
We have to say that, unfortunately, despite the fact that our present 
results extend over almost two orders of magnitudes in terms of the 
correlation length $\xi_L$, we are still not able to clearly distinguish
between the two possibilities, in the sense that both ans{\"a}tze,
Eqs.~\eqref{eq:n2divergent} and~\eqref{eq:n2finite}, return
acceptable values of the $\chi^2/\dof$ test, even if marginally 
better for the convergent ansatz.

\begin{figure}
\centering
\includegraphics[scale=0.52]{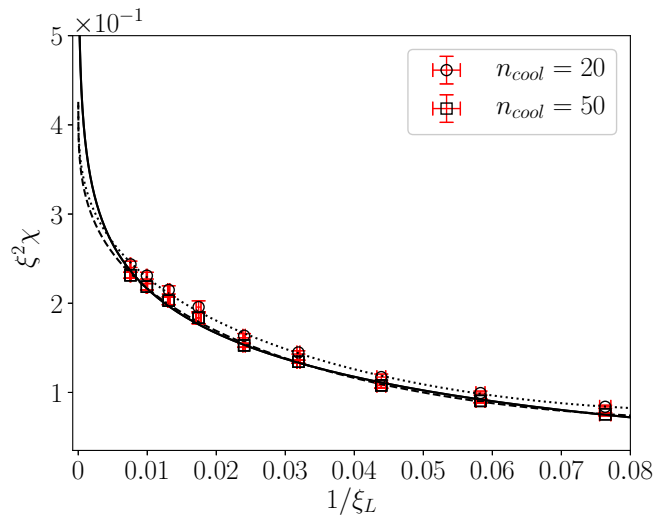}
\caption{Extrapolation towards the continuum limit of $\xi^2 \chi$ for $N=2$ using data in the range $\xi_L> 10$. The solid and dashed lines represent, respectively, the best fits obtained using fit functions $h(x)=a_0^\prime \log(x/a_1^\prime)+a_2^\prime x^2$ (where $x=1/\xi_L$) and $g(x)=a_0+a_1x^2 + a_2x^4 + a_3 x^c$ with measure taken after $n_{cool}=50$ cooling steps. The former best fit yields $\chi^2/\dof = 7.9/6$ while, in the latter case, we obtain the continuum limit $a=0.43(12)$ and the exponent $c=0.30(15)$ ($\chi^2/\dof = 3.8/5$). The dotted line, instead, represents again the best fit obtained using fit function $g(x)=a_0+a_1x^2 + a_2x^4 + a_3 x^c$ but with measures taken after $n_{cool}=20$, which gives $a=0.42(11)$ and $c=0.33(16)$ ($\chi^2/\dof = 3.8/5$).}
\label{fig:continuum_limit_N_2_chi}
\end{figure}

Let us consider, for instance, the data reported in 
Fig.~\ref{fig:continuum_limit_N_2_chi},
where we consider only results obtained for $\xi_L > 10$.
In this range $O(x^4)$ corrections turn out to be unnecessary.
Considering data obtained for $n_{cool} = 50$ we obtain 
$\chi^2/\dof = 7.9/6$ for the divergent 
ansatz in Eq.~\eqref{eq:n2divergent} and 
$\chi^2/\dof = 3.8/5$ for the convergent
ansatz in Eq.~\eqref{eq:n2finite}; in the latter case
we also obtain $c = 0.30(15)$ and 
a prediction $\xi^2 \chi = 0.43(12)$
(for $n_{cool} = 20$, the latter prediction changes 
 to $\xi^2 \chi = 0.42(11)$, $c = 0.33(16)$, 
with  $\chi^2/\dof = 3.8/5$). Notice that, in the 
case of the convergent fit, the coefficient $c$ is at two standard 
deviations from zero, which is the boundary where also this fit becomes 
divergent: this is consistent with the fact that the logarithmic 
fit is marginally acceptable.

The situation does not change appreciably when changing 
the fit range, with both ans{\"a}tze remaining acceptable even if 
with slightly lower values of the 
$\chi^2/\dof$ test in the case of a finite continuum 
susceptibility: if the latter scenario could be assumed a priori
we would conclude $\xi^2 \chi \sim 0.40(15)$ for $N = 2$.
However, the fact that no conclusive answer can be obtained
based on our present data for $N = 2$, is confirmed by looking
at the best fit curves reported in 
Fig.~\ref{fig:continuum_limit_N_2_chi}: 
the two curve profiles (convergent and divergent) 
are hardly distinguishable in the explored range 
of $\xi_L$ and deviate from each other only for 
much larger values of $\xi_L$.
Another independent possibility to discriminate between these two different behaviors is to use data obtained for $N>2$ and extrapolate them towards $N=2$. This topic is covered in Sec.~\ref{sec:small_N_limit_chi}.

\subsection{$N\to2$ limit of $\xi^2\chi$}
\label{sec:small_N_limit_chi}

In this section we aim at tackling the question about
the small $N$ behavior of the topological susceptibility
from another independent front, 
by extrapolating continuum 
results obtained for $\xi^2\chi$ at $N>2$ towards $N=2$.
A summary of such results, including all $N < 10$, 
is reported in Tab.~\ref{tab:chi_vs_N}.
\begin{table}[!t]
\centering
\begin{center}
\begin{tabular}{|c|c|}
\hline
&\\[-1em]
$N$ & $\xi^2 \chi \cdot 10^3$ \\
\hline
&\\[-1em]
3    & 110(5)    \\
4    & 59.5(1.2) \\
5    & 42.90(60) \\
6    & 33.80(30) \\
7    & 27.10(30) \\
8    & 22.40(30) \\
9*   & 20.00(15) \\

\hline
\end{tabular}
\end{center}
\caption{Summary of continuum extrapolations of $\xi^2 \chi$ as a function of $N$. The value marked with * is taken from Ref.~\cite{Bonanno:2018xtd}.}
\label{tab:chi_vs_N}
\end{table}
Using Eq.~\eqref{eq:instanton_size_distribution} and assuming non-interacting instantons, the expected behavior of $\xi^2 \chi$ when approaching the $N\to2$ limit is~\cite{Bonanno:2018xtd}
\beq\label{eq:ansatz_small_N}
\xi^2 \chi \sim \frac{1}{N-2}, \quad N>2.
\eeq
Therefore, we fitted the $N$-dependence of $\xi^2 \chi$ for $N>2$ using the following function:
\beq\label{eq:small_N_fit_function}
F(N)=\frac{a}{(N-N^*)^\gamma}.
\eeq
\begin{figure}[!htb]
\hspace*{-0.3cm}
\includegraphics[scale=0.52]{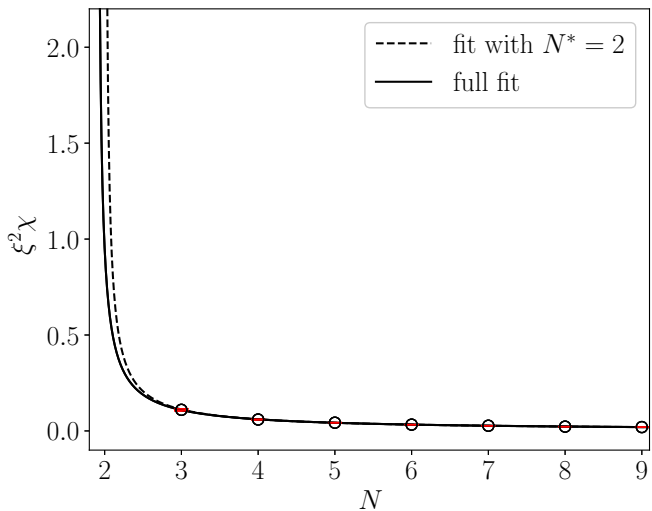}
\caption{Best fit of the small-$N$ behavior of $\xi^2 \chi$ for $N$ ranging from 3 to 9. The solid and dashed lines represent, respectively, the best fits obtained using fit function~\eqref{eq:small_N_fit_function} with $N^* = 2$ or 
left as a free parameter. The best fits yield, respectively, $\chi^2/\dof=5.1/5$ and $\chi^2/\dof=4.56/4$.}
\label{fig:small_N_limit}
\end{figure}

\begin{figure}[!htb]
\hspace*{-0.3cm}
\includegraphics[scale=0.52]{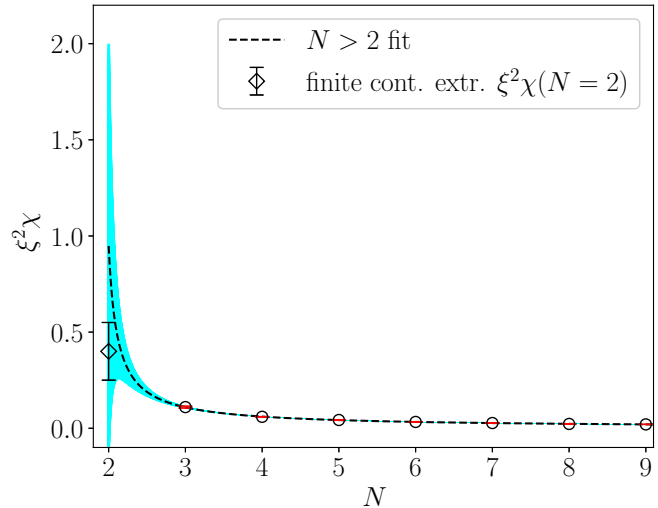}
\caption{Best fit of the small-$N$ behavior of $\xi^2 \chi$ for $N$ ranging from 3 to 9. The dashed line represents the best fit obtained using fit function~\eqref{eq:small_N_fit_function} leaving $N^*$ as a free parameter, the shadowed band represents the fit error and the diamond points represent the hypothetical finite continuum limit obtained from the convergent continuum extrapolation of $\xi^2 \chi$ for $N=2$.}
\label{fig:small_N_limit_N_2_comparison}
\end{figure}

\begin{figure}[!htb]
\hspace*{-0.4cm}
\includegraphics[scale=0.52]{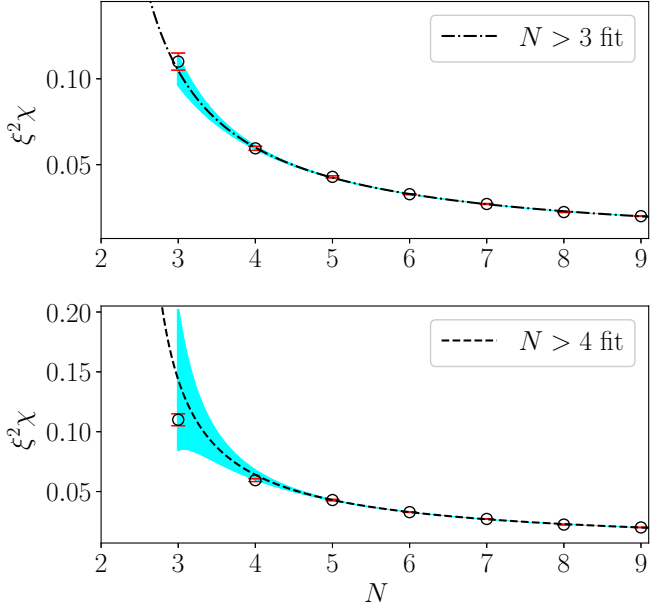}
\caption{Best fits of the small-$N$ behavior of $\xi^2 \chi$ considering just $N>3$ or $N>4$ data. The dashed lines represent the best fits obtained using fit function~\eqref{eq:small_N_fit_function} leaving $N^*$ as a free parameter, while the shadowed bands represent the fit errors. Best fits give, respectively, $N^* = 1.77(31)$ ($\chi^2/\dof = 4.3/3$) and $N^* = 2.40(52)$ ($\chi^2/\dof = 3/2$). Such extrapolations fully support the finite results 
obtained from direct simulations at $N = 3$, both qualitatively and 
quantitatively.}
\label{fig:small_N_limit_N_3_comparison}
\end{figure}

The best fit is quite good and is shown 
in Fig.~\ref{fig:small_N_limit}, the corresponding 
parameters are the following:
\beqnn
a      &=& 0.119(10), \\
N^*    &=& 1.90(14), \\
\gamma &=& 0.91(4), \\
\chi^2/\dof &=& 4.56/4.
\eeqnn
This best fit, yielding a central value for $N^*<2$, technically supports 
a finite extrapolation towards $N=2$; however, $N = 2$ is well within 
one standard deviation from $N^*$, so that even this approach reveals
to be inconclusive, at least for this issue.
As shown in Fig.~\ref{fig:small_N_limit_N_2_comparison}, the error on the best fit blows up as $N\to2$ is approached, being $N^*$ very close to it, 
so that this extrapolation turns out to be compatible 
with the hypothetical finite value found from the convergent 
continuum limit of Sec.~\ref{sec:N_2_results}, 
$\xi^2 \chi = 0.40(15)$, but within 
quite large uncertainties.
As a matter of fact, a best fit of the same data set with~\eqref{eq:ansatz_small_N} but fixing $N^*=2$ yields a perfectly compatible results: $a=0.112(2)$ and $\gamma=0.89(1)$ ($\chi^2/\dof = 5.1/5$); this best fit
is depicted in Fig.~\ref{fig:small_N_limit} as well.
Finally, as a further consistency check of the solidity of these results, we fitted our small-$N$ data also in narrower ranges, in particular 
considering only data for $N>3$ or for $N>4$. Best fits are shown in Fig.~\ref{fig:small_N_limit_N_3_comparison}. In both cases, $N^*$ and $\gamma$ turned out, again, to be compatible with 2 and 1 respectively: 
\begin{align*}
N^* &= 1.77(31), \,  &\gamma  &= 0.94(7),  \quad &N>3,\\
N^* &= 2.40(52), \,  &\gamma  &= 0.82(10), \quad &N>4.
\end{align*}
Moreover, extrapolating these fits towards $N=3$, we obtain values for $\xi^2 \chi(N=3)$ which are, in both cases, in perfect agreement with the continuum limit obtained from direct $N=3$ simulations in Sec.~\ref{sec:N_3_results}.
Based on these results, we conclude that the finiteness of the 
topological susceptibility for $N = 3$ can be definitely assessed.

\subsection{Small-$N$ behavior of the $b_{2n}$ coefficients}
\label{sec:small_N_b2}

In this section we study the small-$N$ behavior of the $b_{2n}$ coefficients, in particular, we aim at checking the hypothesis that these coefficients are finite in the $N\to2$ limit and that, in this limit, they approach the DIGA prediction. With our statistics, we could only obtain reliable estimations of $b_2$, while already $b_4$ always turned out to be compatible with zero,
after continuum extrapolation, in all explored cases. 
Thus, we can presently discuss only the small-$N$ behavior of $b_2$.

\begin{figure}
\hspace*{-0.34cm}
\includegraphics[scale=0.52]{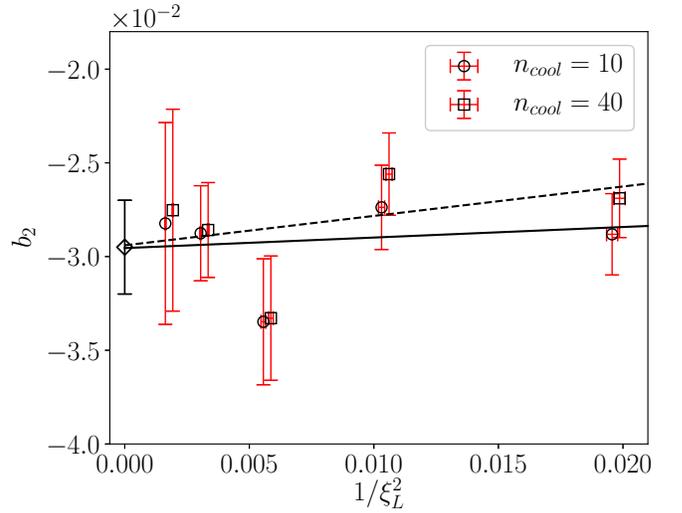}
\caption{Extrapolation towards the continuum limit of $b_2$ for $N=5$ using data in the range $\xi_L>7$. The solid and dashed lines represent the best fits obtained using fit function $f(x)=a_0+a_1x^2$ (where $x=1/\xi_L$) with measures taken, respectively, after $n_{cool}=10$ and $n_{cool}=40$ cooling steps. Best fits yield, respectively, $\chi^2/\dof=2.25/3$ and $\chi^2/\dof=3.24/3$. Data obtained for different numbers of cooling steps have been plotted slightly shifted to improve readability. The diamond point represents our final estimation of the continuum limit.} 
\label{fig:continuum_limit_N_5_b2}
\end{figure}

\begin{figure}
\hspace*{-0.3cm}
\includegraphics[scale=0.52]{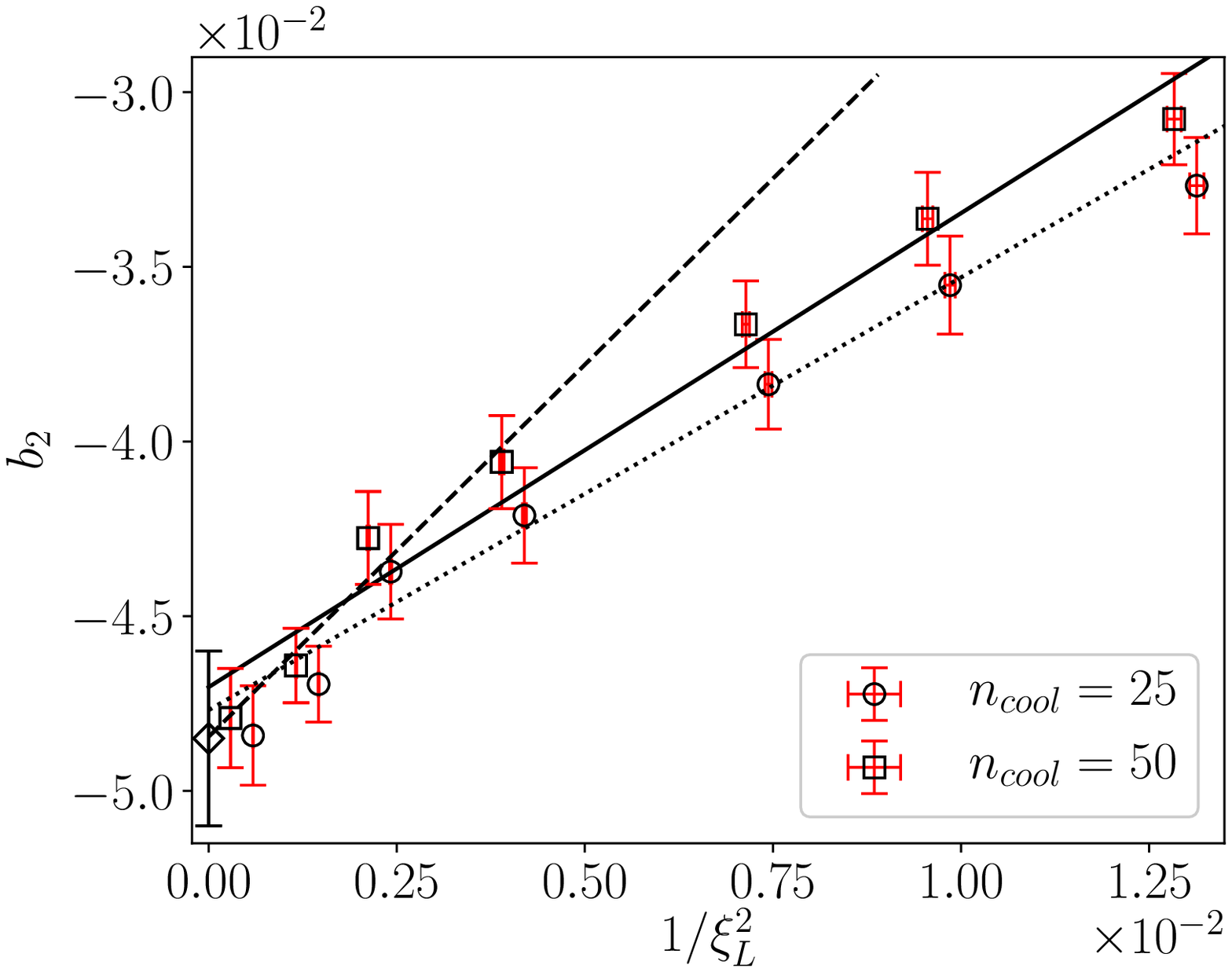}
\caption{Extrapolation towards the continuum limit of $b_2$ for $N=3$. The solid and dashed lines represent the best fits obtained using the fit function $g(x)=a_0+a_1x^2$ (where $x=1/\xi_L$) using measures taken after $n_{cool}=50$ cooling steps in the ranges $\xi_L>8$ and $\xi_L>15$. The best fits yield, respectively, $a_0=-0.0471(8)$ ($\chi^2/\dof=4.65/5$) and $a_0=-0.0485(11)$ ($\chi^2/\dof=1.06/2$). The dotted line represents instead the best fit obtained using measures taken after $n_{cool}=20$ cooling steps in the range $\xi_L>8$. The best fit yields $a_0=-0.0478(8)$ with $\chi^2/\dof=2.85/5$. Data obtained for different numbers of cooling steps have been plotted slightly shifted to improve readability. The diamond point represents our final estimation of the continuum limit.}
\label{fig:continuum_limit_N_3_b2}
\end{figure}

\begin{figure}
\hspace*{-0.35cm}
\includegraphics[scale=0.52]{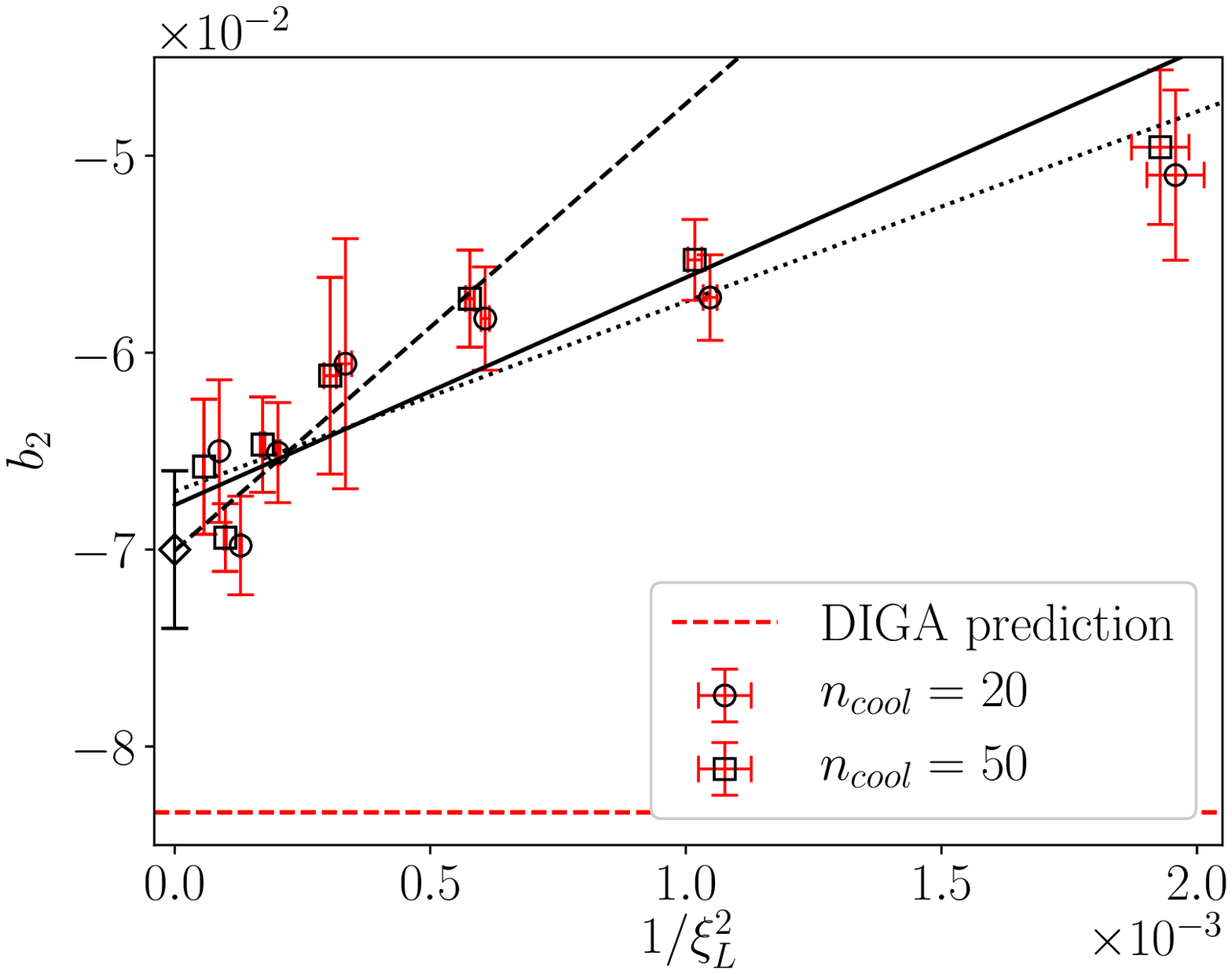}
\caption{Extrapolation towards the continuum limit of $b_2$ for $N=2$. The solid and dashed lines represent the best fits obtained using the fit function $g(x)=a_0+a_1x^2$ (where $x=1/\xi_L$) using measures taken after $n_{cool}=50$ cooling steps in the ranges $\xi_L>20$ and $\xi_L>30$. The best fits yield, respectively, $a_0=-0.0677(13)$ ($\chi^2/\dof=6.95/5$) and $a_0=-0.0700(17)$ ($\chi^2/\dof=2.13/3$). The dotted line represents instead the best fit obtained using measures taken after $n_{cool}=20$ cooling steps in the range $\xi_L>20$. The best fit yields $a_0=-0.0670(16)$ with $\chi^2/\dof=4.55/5$. Data obtained for different numbers of cooling steps have been plotted slightly shifted to improve readability. The diamond point represents our final estimation of the continuum limit.}
\label{fig:continuum_limit_N_2_b2}
\end{figure}

As we have discussed in Sec.~\ref{sec:continuum_limit_small_N}, 
contrary to what happens for the topological susceptibility,
in this case we do not expect a priori modifications to 
the standard form of UV corrections reported in 
Eq.~\eqref{eq:UV_corrections}. As a confirmation of this 
expectation, a continuum extrapolation performed
according to the fit function in Eq.~\eqref{eq:fit_function_N>3} 
turns out to work well for all values 
of $N$, including $N = 2$ and $N = 3$,
with $O(x^4)$ corrections becoming irrelevant and 
not needed when restricting the fit range to 
large enough correlation lengths.
Examples of such extrapolations 
are reported in Figs.~\ref{fig:continuum_limit_N_5_b2},
\ref{fig:continuum_limit_N_3_b2} and
\ref{fig:continuum_limit_N_2_b2}. 

For $N = 2$ and $3$ we have also considered the possible addition
of generic power law corrections to 
Eq.~\eqref{eq:fit_function_N>3} proportional to $x^c$,
as we have done for $\xi^2 \chi$, however this turned out to be 
irrelevant in this case, with modifications to the continuum extrapolation
staying within errors. The only modification which can be noticed in the $N = 2,3$ cases is that the approach to the continuum limit is steeper, however this is 
compensated by the larger values of $\xi_L$ available from our simulations in these cases.

\begin{figure} 
\includegraphics[scale=0.52]{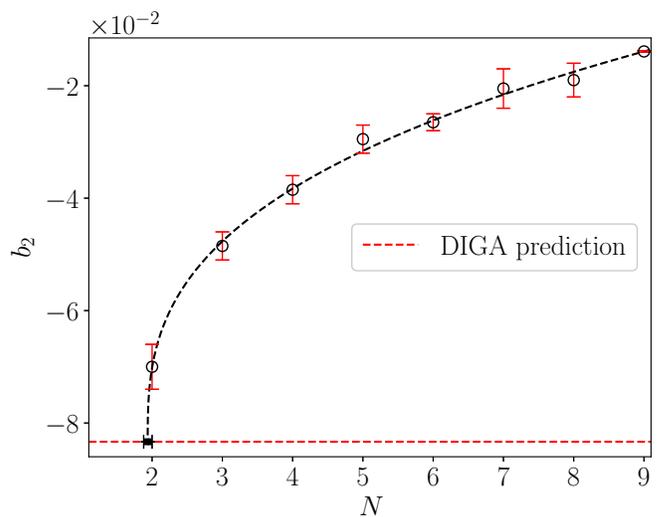}
\caption{Small-$N$ behavior of the quartic coefficient $b_2$. The solid lines represent the best fit obtained using fit function~\eqref{eq:ansatz_small_N_b2}. The best fit yields $N^*=1.94(6)$ with $\chi^2/\dof=1.2/5$. The black point represents the found interval for $N^*$, for which $b_2(N^*)=b_2^\DIGA$.}
\label{fig:small_N_b2}
\end{figure}

All continuum extrapolations are reported in Tab.~\ref{tab:b2_vs_N}.
Also in this case the quoted errors include systematics, which have been
assessed by observing the variation of central fit values when changing 
the fit function, the fit range and the number of cooling steps $n_{cool}$.
\begin{table}
\centering
\begin{center}
\begin{tabular}{|c|c|}
\hline
&\\[-1em]
$N$ & $b_2 \cdot 10^3$ \\
\hline
&\\[-1em]
2  & -70.0(4.0) \\
3  & -48.5(2.5) \\
4  & -38.5(2.5) \\
5  & -29.5(2.5) \\
6  & -28.0(2.0) \\
7  & -20.5(3.5) \\
8  & -19.0(3.0) \\
9* & -13.90(13) \\
\hline
\end{tabular}
\end{center}
\caption{Summary of continuum extrapolations of $b_2$ as a function of $N$. The value marked with * is taken from Ref.~\cite{Bonanno:2018xtd}.}
\label{tab:b2_vs_N}
\end{table}
Our data for $N=2$ confirm that $b_2$ is finite, with 
a continuum estimate $b_2(N=2)=-0.070(4)$ which 
is $\sim 3\sigma$ off from the DIGA prediction $b_2^\DIGA=-1/12\simeq-0.0833$. 
Results reported in 
Ref.~\cite{Bietenholz:2016szu} point also to $b_2 \sim -1/12$, even if 
no continuum limit is performed in that case.\\
We can try to extrapolate, based on present results, the value $N^*$
of $N$
for which $b_2$ reaches the DIGA prediction, to that aim 
we try to fit our data according to a critical behavior like:
\beq\label{eq:ansatz_small_N_b2}
G(N) = b_2^\DIGA + a(N-N^*)^\gamma.
\eeq
While we do not have any argument to support such an ansatz,
it turns out to work quite well: a best fit 
in the range $[2,9]$ 
is reported in Fig.~\ref{fig:small_N_b2} and returns 
the following parameters:
\beqnn
a      &=& 0.0345(18), \\
N^*    &=& 1.94(6), \\
\gamma &=& 0.352(25), \\
\chi^2/{\rm dof} &=& 1.2/5.
\eeqnn
It is interesting to observe that the $N^*$ found in this way from 
the analysis of $b_2$ is perfectly compatible with the one found for the critical fit of $\xi^2 \chi$, 
and still compatible with $N^* = 2$.

\section{Conclusions}
\label{sec:final}

In this work, we have presented a systematic numerical study of 
the peculiar features of the $\theta$-dependence around $\theta=0$ of the vacuum energy $f(\theta)$ of $2d$ $CP^{N-1}$ models in the small-$N$ limit.
To that aim, we have performed numerical simulations for $N \in [2,8]$.

One of the most interesting questions regards the value 
of $N$, if any, at which the topological susceptibility $\chi$ 
diverges: this is predicted to be $N = 2$
by perturbative computations of the instanton size distribution.
A  second interesting question regards the fate of the 
$b_{2n}$ coefficients, which parametrize the 
non-quadratic part of $\theta$-dependence and which 
could possibly approach the values predicted by the 
Dilute Instanton Gas Approximation at the point where 
$\chi$ diverges, if the theory can be approximated in this 
case by a gas of small and 
non-interacting instantons and anti-instantons.

Our strategy has been twofold. On one hand, we have dedicated
particular efforts to correctly assess the continuum limit of 
$\xi^2 \chi$ for $N = 2$ and $N = 3$ from direct simulations
of these theories: to that aim, we have performed
simulations on lattices with correlations lengths ranging from a few units
up to $O(10^2)$. On the other hand, we have exploited results
obtained for larger values of $N$, where the continuum
extrapolation is easier, in order to perform a small-$N$ extrapolation.
Based on this double strategy, we have obtained consistent and conclusive 
evidence that the topological susceptibility is finite for 
$N = 3$, providing the estimate $\xi^2 \chi = 0.110(5)$.
We would like to stress that, since the convergence of $\xi^2 \chi$ 
in the continuum limit for $N = 3$ was debated in previous literature,
a double check with two different and independent methods is 
important for our final assessment about this issue.

On the other hand, results for $N = 2$ are still inconclusive:
results obtained directly at $N = 2$ 
are consistent with a logarithmically-divergent 
continuum extrapolation, but do not yet exclude 
a finite continuum value with $\xi^2 \chi \sim 0.4$,
which is even marginally favored from the point of the
$\chi^2/\dof$ test. 
A similar picture emerges from the extrapolation from 
results obtained for $N > 2$, which provides evidence for a 
critical behavior $\xi^2 \chi \propto 1/(N - N^*)^\gamma$, with
$N^* = 1.90(14)$. Therefore, future numerical studies
are still needed in this case to definitely settle the issue.

As for the $b_2$ coefficient, we have provided continuum 
extrapolations down to $N = 2$.
While there is no compelling reason to expect that the 
DIGA prediction be valid at the point where $\xi^2 \chi$ 
diverges, it is interesting to observe that our numerical 
results are consistent with that. 
For $N = 2$ we obtain
$b_2(N=2)=-0.070(4)$, i.e., around $3 \sigma$ off 
the DIGA value $b_2^\DIGA=-1/12\simeq -0.0833$,
while an extrapolation from our results at all values of $N$
(see Eq.~\eqref{eq:ansatz_small_N_b2})
provides evidence for $b_2$ reaching $b_2^\DIGA$ for 
$N = 1.94(6)$, which is consistent with the 
value $N^*$ at which $\xi^2 \chi$ diverges reported above.

\acknowledgments
The authors thank C.~Bonati and T.~Sulejmanpasic for useful discussions. Numerical simulations have been performed at the Scientific Computing Center at INFN-PISA and on the MARCONI machine at CINECA, based on the agreement between INFN and CINECA (under projects INF19\_npqcd and INF20\_npqcd).


\begin{thebibliography}{58}
	\expandafter\ifx\csname natexlab\endcsname\relax\def\natexlab#1{#1}\fi
	\expandafter\ifx\csname bibnamefont\endcsname\relax
	\def\bibnamefont#1{#1}\fi
	\expandafter\ifx\csname bibfnamefont\endcsname\relax
	\def\bibfnamefont#1{#1}\fi
	\expandafter\ifx\csname citenamefont\endcsname\relax
	\def\citenamefont#1{#1}\fi
	\expandafter\ifx\csname url\endcsname\relax
	\def\url#1{\texttt{#1}}\fi
	\expandafter\ifx\csname urlprefix\endcsname\relax\def\urlprefix{URL }\fi
	\providecommand{\bibinfo}[2]{#2}
	\providecommand{\eprint}[2][]{\url{#2}}
	
	\bibitem[{\citenamefont{D'Adda et~al.}(1978)\citenamefont{D'Adda, L{\"u}scher,
			and Di~Vecchia}}]{DAdda:1978vbw}
	\bibinfo{author}{\bibfnamefont{A.}~\bibnamefont{D'Adda}},
	\bibinfo{author}{\bibfnamefont{M.}~\bibnamefont{L{\"u}scher}},
	\bibnamefont{and}
	\bibinfo{author}{\bibfnamefont{P.}~\bibnamefont{Di~Vecchia}},
	\bibinfo{journal}{Nucl. Phys. B} \textbf{\bibinfo{volume}{146}},
	\bibinfo{pages}{63} (\bibinfo{year}{1978}).
	
	\bibitem[{\citenamefont{Shifman}(2012)}]{advanced_topics_QFT}
	\bibinfo{author}{\bibfnamefont{M.}~\bibnamefont{Shifman}},
	\emph{\bibinfo{title}{Advanced topics in {Q}uantum {F}ield {T}heory}}
	(\bibinfo{publisher}{Cambridge University Press},
	\bibinfo{address}{Cambridge}, \bibinfo{year}{2012}), pp.
	\bibinfo{pages}{171--268, 361--367}.
	
	\bibitem[{\citenamefont{Vicari and Panagopoulos}(2009)}]{Vicari:2008jw}
	\bibinfo{author}{\bibfnamefont{E.}~\bibnamefont{Vicari}} \bibnamefont{and}
	\bibinfo{author}{\bibfnamefont{H.}~\bibnamefont{Panagopoulos}},
	\bibinfo{journal}{Phys. Rept.} \textbf{\bibinfo{volume}{470}},
	\bibinfo{pages}{93} (\bibinfo{year}{2009}), \eprint{0803.1593}.
	
	\bibitem[{\citenamefont{Witten}(1980)}]{Witten:1980sp}
	\bibinfo{author}{\bibfnamefont{E.}~\bibnamefont{Witten}},
	\bibinfo{journal}{Annals Phys.} \textbf{\bibinfo{volume}{128}},
	\bibinfo{pages}{363} (\bibinfo{year}{1980}).
	
	\bibitem[{\citenamefont{Witten}(1998)}]{Witten:1998uka}
	\bibinfo{author}{\bibfnamefont{E.}~\bibnamefont{Witten}},
	\bibinfo{journal}{Phys. Rev. Lett.} \textbf{\bibinfo{volume}{81}},
	\bibinfo{pages}{2862} (\bibinfo{year}{1998}), \eprint{hep-th/9807109}.
	
	\bibitem[{\citenamefont{Rossi}(2016)}]{Rossi:2016uce}
	\bibinfo{author}{\bibfnamefont{P.}~\bibnamefont{Rossi}},
	\bibinfo{journal}{Phys. Rev.} \textbf{\bibinfo{volume}{D94}},
	\bibinfo{pages}{045013} (\bibinfo{year}{2016}), \eprint{1606.07252}.
	
	\bibitem[{\citenamefont{Bonati et~al.}(2016{\natexlab{a}})\citenamefont{Bonati,
			D'Elia, Rossi, and Vicari}}]{Bonati:2016tvi}
	\bibinfo{author}{\bibfnamefont{C.}~\bibnamefont{Bonati}},
	\bibinfo{author}{\bibfnamefont{M.}~\bibnamefont{D'Elia}},
	\bibinfo{author}{\bibfnamefont{P.}~\bibnamefont{Rossi}}, \bibnamefont{and}
	\bibinfo{author}{\bibfnamefont{E.}~\bibnamefont{Vicari}},
	\bibinfo{journal}{Phys. Rev.} \textbf{\bibinfo{volume}{D94}},
	\bibinfo{pages}{085017} (\bibinfo{year}{2016}{\natexlab{a}}),
	\eprint{1607.06360}.
	
	\bibitem[{\citenamefont{Berni et~al.}(2019)\citenamefont{Berni, Bonanno, and
			D'Elia}}]{Berni:2019bch}
	\bibinfo{author}{\bibfnamefont{M.}~\bibnamefont{Berni}},
	\bibinfo{author}{\bibfnamefont{C.}~\bibnamefont{Bonanno}}, \bibnamefont{and}
	\bibinfo{author}{\bibfnamefont{M.}~\bibnamefont{D'Elia}},
	\bibinfo{journal}{Phys. Rev.} \textbf{\bibinfo{volume}{D100}},
	\bibinfo{pages}{114509} (\bibinfo{year}{2019}), \eprint{1911.03384}.
	
	\bibitem[{\citenamefont{Del~Debbio et~al.}(2006)\citenamefont{Del~Debbio,
			Manca, Panagopoulos, Skouroupathis, and Vicari}}]{DelDebbio:2006yuf}
	\bibinfo{author}{\bibfnamefont{L.}~\bibnamefont{Del~Debbio}},
	\bibinfo{author}{\bibfnamefont{G.~M.} \bibnamefont{Manca}},
	\bibinfo{author}{\bibfnamefont{H.}~\bibnamefont{Panagopoulos}},
	\bibinfo{author}{\bibfnamefont{A.}~\bibnamefont{Skouroupathis}},
	\bibnamefont{and} \bibinfo{author}{\bibfnamefont{E.}~\bibnamefont{Vicari}},
	\bibinfo{journal}{JHEP} \textbf{\bibinfo{volume}{06}}, \bibinfo{pages}{005}
	(\bibinfo{year}{2006}), \eprint{hep-th/0603041}.
	
	\bibitem[{\citenamefont{Campostrini and Rossi}(1991)}]{Campostrini:1991kv}
	\bibinfo{author}{\bibfnamefont{M.}~\bibnamefont{Campostrini}} \bibnamefont{and}
	\bibinfo{author}{\bibfnamefont{P.}~\bibnamefont{Rossi}},
	\bibinfo{journal}{Phys. Lett. B} \textbf{\bibinfo{volume}{272}},
	\bibinfo{pages}{305} (\bibinfo{year}{1991}).
	
	\bibitem[{\citenamefont{Witten}(1979)}]{Witten:1978bc}
	\bibinfo{author}{\bibfnamefont{E.}~\bibnamefont{Witten}},
	\bibinfo{journal}{Nucl. Phys. B} \textbf{\bibinfo{volume}{149}},
	\bibinfo{pages}{285} (\bibinfo{year}{1979}).
	
	\bibitem[{\citenamefont{Veneziano}(1979)}]{Veneziano:1979ec}
	\bibinfo{author}{\bibfnamefont{G.}~\bibnamefont{Veneziano}},
	\bibinfo{journal}{Nucl. Phys. B} \textbf{\bibinfo{volume}{159}},
	\bibinfo{pages}{213} (\bibinfo{year}{1979}).
	
	\bibitem[{\citenamefont{Bonanno et~al.}(2019)\citenamefont{Bonanno, Bonati, and
			D'Elia}}]{Bonanno:2018xtd}
	\bibinfo{author}{\bibfnamefont{C.}~\bibnamefont{Bonanno}},
	\bibinfo{author}{\bibfnamefont{C.}~\bibnamefont{Bonati}}, \bibnamefont{and}
	\bibinfo{author}{\bibfnamefont{M.}~\bibnamefont{D'Elia}},
	\bibinfo{journal}{JHEP} \textbf{\bibinfo{volume}{01}}, \bibinfo{pages}{003}
	(\bibinfo{year}{2019}), \eprint{1807.11357}.
	
	\bibitem[{\citenamefont{Jevicki}(1977)}]{Jevicki:1977yd}
	\bibinfo{author}{\bibfnamefont{A.}~\bibnamefont{Jevicki}},
	\bibinfo{journal}{Nucl. Phys.} \textbf{\bibinfo{volume}{B127}},
	\bibinfo{pages}{125} (\bibinfo{year}{1977}).
	
	\bibitem[{\citenamefont{Forster}(1977)}]{Forster:1977jv}
	\bibinfo{author}{\bibfnamefont{D.}~\bibnamefont{Forster}},
	\bibinfo{journal}{Nucl. Phys.} \textbf{\bibinfo{volume}{B130}},
	\bibinfo{pages}{38} (\bibinfo{year}{1977}).
	
	\bibitem[{\citenamefont{Berg and L{\"u}scher}(1979)}]{Berg:1979uq}
	\bibinfo{author}{\bibfnamefont{B.}~\bibnamefont{Berg}} \bibnamefont{and}
	\bibinfo{author}{\bibfnamefont{M.}~\bibnamefont{L{\"u}scher}},
	\bibinfo{journal}{Commun. Math. Phys.} \textbf{\bibinfo{volume}{69}},
	\bibinfo{pages}{57} (\bibinfo{year}{1979}).
	
	\bibitem[{\citenamefont{Fateev et~al.}(1979)\citenamefont{Fateev, Frolov, and
			Shvarts}}]{Fateev:1979dc}
	\bibinfo{author}{\bibfnamefont{V.~A.} \bibnamefont{Fateev}},
	\bibinfo{author}{\bibfnamefont{I.~V.} \bibnamefont{Frolov}},
	\bibnamefont{and} \bibinfo{author}{\bibfnamefont{A.~S.}
		\bibnamefont{Shvarts}}, \bibinfo{journal}{Nucl. Phys.}
	\textbf{\bibinfo{volume}{B154}}, \bibinfo{pages}{1} (\bibinfo{year}{1979}).
	
	\bibitem[{\citenamefont{Richard and Rouet}(1983)}]{Richard:1981bv}
	\bibinfo{author}{\bibfnamefont{J.-L.} \bibnamefont{Richard}} \bibnamefont{and}
	\bibinfo{author}{\bibfnamefont{A.}~\bibnamefont{Rouet}},
	\bibinfo{journal}{Nucl. Phys.} \textbf{\bibinfo{volume}{B211}},
	\bibinfo{pages}{447} (\bibinfo{year}{1983}).
	
	\bibitem[{\citenamefont{D'Elia et~al.}(1995)\citenamefont{D'Elia, Farchioni,
			and Papa}}]{DElia:1995zja}
	\bibinfo{author}{\bibfnamefont{M.}~\bibnamefont{D'Elia}},
	\bibinfo{author}{\bibfnamefont{F.}~\bibnamefont{Farchioni}},
	\bibnamefont{and} \bibinfo{author}{\bibfnamefont{A.}~\bibnamefont{Papa}},
	\bibinfo{journal}{Nucl. Phys. B} \textbf{\bibinfo{volume}{456}},
	\bibinfo{pages}{313} (\bibinfo{year}{1995}), \eprint{hep-lat/9505004}.
	
	\bibitem[{\citenamefont{D'Elia et~al.}(1997)\citenamefont{D'Elia, Farchioni,
			and Papa}}]{DElia:1995wxi}
	\bibinfo{author}{\bibfnamefont{M.}~\bibnamefont{D'Elia}},
	\bibinfo{author}{\bibfnamefont{F.}~\bibnamefont{Farchioni}},
	\bibnamefont{and} \bibinfo{author}{\bibfnamefont{A.}~\bibnamefont{Papa}},
	\bibinfo{journal}{Phys. Rev.} \textbf{\bibinfo{volume}{D55}},
	\bibinfo{pages}{2274} (\bibinfo{year}{1997}), \eprint{hep-lat/9511021}.
	
	\bibitem[{\citenamefont{Blatter et~al.}(1996)\citenamefont{Blatter, Burkhalter,
			Hasenfratz, and Niedermayer}}]{Blatter:1995ik}
	\bibinfo{author}{\bibfnamefont{M.}~\bibnamefont{Blatter}},
	\bibinfo{author}{\bibfnamefont{R.}~\bibnamefont{Burkhalter}},
	\bibinfo{author}{\bibfnamefont{P.}~\bibnamefont{Hasenfratz}},
	\bibnamefont{and}
	\bibinfo{author}{\bibfnamefont{F.}~\bibnamefont{Niedermayer}},
	\bibinfo{journal}{Phys. Rev.} \textbf{\bibinfo{volume}{D53}},
	\bibinfo{pages}{923} (\bibinfo{year}{1996}), \eprint{hep-lat/9508028}.
	
	\bibitem[{\citenamefont{Ahmad et~al.}(2005)\citenamefont{Ahmad, Lenaghan, and
			Thacker}}]{Ahmad:2005dr}
	\bibinfo{author}{\bibfnamefont{S.}~\bibnamefont{Ahmad}},
	\bibinfo{author}{\bibfnamefont{J.~T.} \bibnamefont{Lenaghan}},
	\bibnamefont{and} \bibinfo{author}{\bibfnamefont{H.~B.}
		\bibnamefont{Thacker}}, \bibinfo{journal}{Phys. Rev. D}
	\textbf{\bibinfo{volume}{72}}, \bibinfo{pages}{114511}
	(\bibinfo{year}{2005}), \eprint{hep-lat/0509066}.
	
	\bibitem[{\citenamefont{Lian and Thacker}(2007)}]{Lian:2006ky}
	\bibinfo{author}{\bibfnamefont{Y.}~\bibnamefont{Lian}} \bibnamefont{and}
	\bibinfo{author}{\bibfnamefont{H.~B.} \bibnamefont{Thacker}},
	\bibinfo{journal}{Phys. Rev.} \textbf{\bibinfo{volume}{D75}},
	\bibinfo{pages}{065031} (\bibinfo{year}{2007}), \eprint{hep-lat/0607026}.
	
	\bibitem[{\citenamefont{Bietenholz et~al.}(2010)\citenamefont{Bietenholz,
			Gerber, Pepe, and Wiese}}]{Bietenholz:2010xg}
	\bibinfo{author}{\bibfnamefont{W.}~\bibnamefont{Bietenholz}},
	\bibinfo{author}{\bibfnamefont{U.}~\bibnamefont{Gerber}},
	\bibinfo{author}{\bibfnamefont{M.}~\bibnamefont{Pepe}}, \bibnamefont{and}
	\bibinfo{author}{\bibfnamefont{U.-J.} \bibnamefont{Wiese}},
	\bibinfo{journal}{JHEP} \textbf{\bibinfo{volume}{12}}, \bibinfo{pages}{020}
	(\bibinfo{year}{2010}), \eprint{1009.2146}.
	
	\bibitem[{\citenamefont{Bietenholz et~al.}(2018)\citenamefont{Bietenholz,
			de~Forcrand, Gerber, Mej{\'i}a-D{\'i}az, and Sandoval}}]{Bietenholz:2018agd}
	\bibinfo{author}{\bibfnamefont{W.}~\bibnamefont{Bietenholz}},
	\bibinfo{author}{\bibfnamefont{P.}~\bibnamefont{de~Forcrand}},
	\bibinfo{author}{\bibfnamefont{U.}~\bibnamefont{Gerber}},
	\bibinfo{author}{\bibfnamefont{H.}~\bibnamefont{Mej{\'i}a-D{\'i}az}},
	\bibnamefont{and} \bibinfo{author}{\bibfnamefont{I.~O.}
		\bibnamefont{Sandoval}}, \bibinfo{journal}{Phys. Rev. D}
	\textbf{\bibinfo{volume}{98}}, \bibinfo{pages}{114501}
	(\bibinfo{year}{2018}), \eprint{1808.08129}.
	
	\bibitem[{\citenamefont{Petcher and L{\"u}scher}(1983)}]{Petcher198353}
	\bibinfo{author}{\bibfnamefont{D.}~\bibnamefont{Petcher}} \bibnamefont{and}
	\bibinfo{author}{\bibfnamefont{M.}~\bibnamefont{L{\"u}scher}},
	\bibinfo{journal}{Nuclear Physics B} \textbf{\bibinfo{volume}{225}},
	\bibinfo{pages}{53 } (\bibinfo{year}{1983}), ISSN \bibinfo{issn}{0550-3213}.
	
	\bibitem[{\citenamefont{Gaiotto et~al.}(2017)\citenamefont{Gaiotto, Kapustin,
			Komargodski, and Seiberg}}]{Gaiotto:2017yup}
	\bibinfo{author}{\bibfnamefont{D.}~\bibnamefont{Gaiotto}},
	\bibinfo{author}{\bibfnamefont{A.}~\bibnamefont{Kapustin}},
	\bibinfo{author}{\bibfnamefont{Z.}~\bibnamefont{Komargodski}},
	\bibnamefont{and} \bibinfo{author}{\bibfnamefont{N.}~\bibnamefont{Seiberg}},
	\bibinfo{journal}{JHEP} \textbf{\bibinfo{volume}{05}}, \bibinfo{pages}{091}
	(\bibinfo{year}{2017}), \eprint{1703.00501}.
	
	\bibitem[{\citenamefont{Sulejmanpasic and
			Tanizaki}(2018)}]{Sulejmanpasic:2018upi}
	\bibinfo{author}{\bibfnamefont{T.}~\bibnamefont{Sulejmanpasic}}
	\bibnamefont{and} \bibinfo{author}{\bibfnamefont{Y.}~\bibnamefont{Tanizaki}},
	\bibinfo{journal}{Phys. Rev. B} \textbf{\bibinfo{volume}{97}},
	\bibinfo{pages}{144201} (\bibinfo{year}{2018}), \eprint{1802.02153}.
	
	\bibitem[{\citenamefont{Affleck and Haldane}(1987)}]{Affleck:1987ch}
	\bibinfo{author}{\bibfnamefont{I.}~\bibnamefont{Affleck}} \bibnamefont{and}
	\bibinfo{author}{\bibfnamefont{F.~D.~M.} \bibnamefont{Haldane}},
	\bibinfo{journal}{Phys. Rev. B} \textbf{\bibinfo{volume}{36}},
	\bibinfo{pages}{5291} (\bibinfo{year}{1987}).
	
	\bibitem[{\citenamefont{Affleck}(1991)}]{Affleck:1991tj}
	\bibinfo{author}{\bibfnamefont{I.}~\bibnamefont{Affleck}},
	\bibinfo{journal}{Phys. Rev. Lett.} \textbf{\bibinfo{volume}{66}},
	\bibinfo{pages}{2429} (\bibinfo{year}{1991}).
	
	\bibitem[{\citenamefont{Alles and Papa}(2008)}]{Alles:2007br}
	\bibinfo{author}{\bibfnamefont{B.}~\bibnamefont{Alles}} \bibnamefont{and}
	\bibinfo{author}{\bibfnamefont{A.}~\bibnamefont{Papa}},
	\bibinfo{journal}{Phys. Rev. D} \textbf{\bibinfo{volume}{77}},
	\bibinfo{pages}{056008} (\bibinfo{year}{2008}), \eprint{0711.1496}.
	
	\bibitem[{\citenamefont{Sulejmanpasic et~al.}(2020)\citenamefont{Sulejmanpasic,
			G\"oschl, and Gattringer}}]{Sulejmanpasic:2020lyq}
	\bibinfo{author}{\bibfnamefont{T.}~\bibnamefont{Sulejmanpasic}},
	\bibinfo{author}{\bibfnamefont{D.}~\bibnamefont{G\"oschl}}, \bibnamefont{and}
	\bibinfo{author}{\bibfnamefont{C.}~\bibnamefont{Gattringer}},
	\bibinfo{journal}{Phys. Rev. Lett.} \textbf{\bibinfo{volume}{125}},
	\bibinfo{pages}{201602} (\bibinfo{year}{2020}), \eprint{2007.06323}.
	
	\bibitem[{\citenamefont{Bietenholz et~al.}(2016)\citenamefont{Bietenholz,
			Cichy, de~Forcrand, Dromard, and Gerber}}]{Bietenholz:2016szu}
	\bibinfo{author}{\bibfnamefont{W.}~\bibnamefont{Bietenholz}},
	\bibinfo{author}{\bibfnamefont{K.}~\bibnamefont{Cichy}},
	\bibinfo{author}{\bibfnamefont{P.}~\bibnamefont{de~Forcrand}},
	\bibinfo{author}{\bibfnamefont{A.}~\bibnamefont{Dromard}}, \bibnamefont{and}
	\bibinfo{author}{\bibfnamefont{U.}~\bibnamefont{Gerber}}
	(\bibinfo{year}{2016}), vol. \bibinfo{volume}{LATTICE2016}, p.
	\bibinfo{pages}{321}, \eprint{1610.00685}.
	
	\bibitem[{\citenamefont{Campostrini
			et~al.}(1992{\natexlab{a}})\citenamefont{Campostrini, Rossi, and
			Vicari}}]{Campostrini:1992ar}
	\bibinfo{author}{\bibfnamefont{M.}~\bibnamefont{Campostrini}},
	\bibinfo{author}{\bibfnamefont{P.}~\bibnamefont{Rossi}}, \bibnamefont{and}
	\bibinfo{author}{\bibfnamefont{E.}~\bibnamefont{Vicari}},
	\bibinfo{journal}{Phys. Rev. D} \textbf{\bibinfo{volume}{46}},
	\bibinfo{pages}{2647} (\bibinfo{year}{1992}{\natexlab{a}}).
	
	\bibitem[{\citenamefont{Caracciolo and Pelissetto}(1998)}]{Caracciolo:1998gga}
	\bibinfo{author}{\bibfnamefont{S.}~\bibnamefont{Caracciolo}} \bibnamefont{and}
	\bibinfo{author}{\bibfnamefont{A.}~\bibnamefont{Pelissetto}},
	\bibinfo{journal}{Phys. Rev. D} \textbf{\bibinfo{volume}{58}},
	\bibinfo{pages}{105007} (\bibinfo{year}{1998}), \eprint{hep-lat/9804001}.
	
	\bibitem[{\citenamefont{Campostrini et~al.}(1988)\citenamefont{Campostrini,
			Di~Giacomo, and Panagopoulos}}]{Campostrini:1988ab}
	\bibinfo{author}{\bibfnamefont{M.}~\bibnamefont{Campostrini}},
	\bibinfo{author}{\bibfnamefont{A.}~\bibnamefont{Di~Giacomo}},
	\bibnamefont{and}
	\bibinfo{author}{\bibfnamefont{H.}~\bibnamefont{Panagopoulos}},
	\bibinfo{journal}{Phys. Lett.} \textbf{\bibinfo{volume}{B212}},
	\bibinfo{pages}{206} (\bibinfo{year}{1988}).
	
	\bibitem[{\citenamefont{Farchioni and Papa}(1993)}]{Farchioni:1993jd}
	\bibinfo{author}{\bibfnamefont{F.}~\bibnamefont{Farchioni}} \bibnamefont{and}
	\bibinfo{author}{\bibfnamefont{A.}~\bibnamefont{Papa}},
	\bibinfo{journal}{Phys. Lett. B} \textbf{\bibinfo{volume}{306}},
	\bibinfo{pages}{108} (\bibinfo{year}{1993}).
	
	\bibitem[{\citenamefont{Berg and L{\"u}scher}(1981)}]{Berg:1981er}
	\bibinfo{author}{\bibfnamefont{B.}~\bibnamefont{Berg}} \bibnamefont{and}
	\bibinfo{author}{\bibfnamefont{M.}~\bibnamefont{L{\"u}scher}},
	\bibinfo{journal}{Nucl. Phys. B} \textbf{\bibinfo{volume}{190}},
	\bibinfo{pages}{412} (\bibinfo{year}{1981}).
	
	\bibitem[{\citenamefont{Berg}(1981)}]{Berg:1981nw}
	\bibinfo{author}{\bibfnamefont{B.}~\bibnamefont{Berg}}, \bibinfo{journal}{Phys.
		Lett. B} \textbf{\bibinfo{volume}{104}}, \bibinfo{pages}{475}
	(\bibinfo{year}{1981}).
	
	\bibitem[{\citenamefont{Campostrini
			et~al.}(1992{\natexlab{b}})\citenamefont{Campostrini, Rossi, and
			Vicari}}]{Campostrini:1992it}
	\bibinfo{author}{\bibfnamefont{M.}~\bibnamefont{Campostrini}},
	\bibinfo{author}{\bibfnamefont{P.}~\bibnamefont{Rossi}}, \bibnamefont{and}
	\bibinfo{author}{\bibfnamefont{E.}~\bibnamefont{Vicari}},
	\bibinfo{journal}{Phys. Rev. D} \textbf{\bibinfo{volume}{46}},
	\bibinfo{pages}{4643} (\bibinfo{year}{1992}{\natexlab{b}}),
	\eprint{hep-lat/9207032}.
	
	\bibitem[{\citenamefont{D'Elia}(2003)}]{DElia:2003zne}
	\bibinfo{author}{\bibfnamefont{M.}~\bibnamefont{D'Elia}},
	\bibinfo{journal}{Nucl. Phys.} \textbf{\bibinfo{volume}{B661}},
	\bibinfo{pages}{139} (\bibinfo{year}{2003}), \eprint{hep-lat/0302007}.
	
	\bibitem[{\citenamefont{Iwasaki and Yoshie}(1983)}]{Iwasaki:1983bv}
	\bibinfo{author}{\bibfnamefont{Y.}~\bibnamefont{Iwasaki}} \bibnamefont{and}
	\bibinfo{author}{\bibfnamefont{T.}~\bibnamefont{Yoshie}},
	\bibinfo{journal}{Phys. Lett. B} \textbf{\bibinfo{volume}{131}},
	\bibinfo{pages}{159} (\bibinfo{year}{1983}).
	
	\bibitem[{\citenamefont{Itoh et~al.}(1984)\citenamefont{Itoh, Iwasaki, and
			Yoshie}}]{Itoh:1984pr}
	\bibinfo{author}{\bibfnamefont{S.}~\bibnamefont{Itoh}},
	\bibinfo{author}{\bibfnamefont{Y.}~\bibnamefont{Iwasaki}}, \bibnamefont{and}
	\bibinfo{author}{\bibfnamefont{T.}~\bibnamefont{Yoshie}},
	\bibinfo{journal}{Phys. Lett. B} \textbf{\bibinfo{volume}{147}},
	\bibinfo{pages}{141} (\bibinfo{year}{1984}).
	
	\bibitem[{\citenamefont{Teper}(1985)}]{Teper:1985rb}
	\bibinfo{author}{\bibfnamefont{M.}~\bibnamefont{Teper}},
	\bibinfo{journal}{Phys. Lett. B} \textbf{\bibinfo{volume}{162}},
	\bibinfo{pages}{357} (\bibinfo{year}{1985}).
	
	\bibitem[{\citenamefont{Ilgenfritz et~al.}(1986)\citenamefont{Ilgenfritz,
			Laursen, Schierholz, M{\"u}ller-Preussker, and Schiller}}]{Ilgenfritz:1985dz}
	\bibinfo{author}{\bibfnamefont{E.-M.} \bibnamefont{Ilgenfritz}},
	\bibinfo{author}{\bibfnamefont{M.}~\bibnamefont{Laursen}},
	\bibinfo{author}{\bibfnamefont{G.}~\bibnamefont{Schierholz}},
	\bibinfo{author}{\bibfnamefont{M.}~\bibnamefont{M{\"u}ller-Preussker}},
	\bibnamefont{and} \bibinfo{author}{\bibfnamefont{H.}~\bibnamefont{Schiller}},
	\bibinfo{journal}{Nucl. Phys. B} \textbf{\bibinfo{volume}{268}},
	\bibinfo{pages}{693} (\bibinfo{year}{1986}).
	
	\bibitem[{\citenamefont{Campostrini et~al.}(1990)\citenamefont{Campostrini,
			Di~Giacomo, Panagopoulos, and Vicari}}]{Campostrini:1989dh}
	\bibinfo{author}{\bibfnamefont{M.}~\bibnamefont{Campostrini}},
	\bibinfo{author}{\bibfnamefont{A.}~\bibnamefont{Di~Giacomo}},
	\bibinfo{author}{\bibfnamefont{H.}~\bibnamefont{Panagopoulos}},
	\bibnamefont{and} \bibinfo{author}{\bibfnamefont{E.}~\bibnamefont{Vicari}},
	\bibinfo{journal}{Nucl. Phys. B} \textbf{\bibinfo{volume}{329}},
	\bibinfo{pages}{683} (\bibinfo{year}{1990}).
	
	\bibitem[{\citenamefont{Alles et~al.}(2000)\citenamefont{Alles, Cosmai, D'Elia,
			and Papa}}]{Alles:2000sc}
	\bibinfo{author}{\bibfnamefont{B.}~\bibnamefont{Alles}},
	\bibinfo{author}{\bibfnamefont{L.}~\bibnamefont{Cosmai}},
	\bibinfo{author}{\bibfnamefont{M.}~\bibnamefont{D'Elia}}, \bibnamefont{and}
	\bibinfo{author}{\bibfnamefont{A.}~\bibnamefont{Papa}},
	\bibinfo{journal}{Phys. Rev. D} \textbf{\bibinfo{volume}{62}},
	\bibinfo{pages}{094507} (\bibinfo{year}{2000}), \eprint{hep-lat/0001027}.
	
	\bibitem[{\citenamefont{L{\"u}scher}(2010{\natexlab{a}})}]{Luscher:2009eq}
	\bibinfo{author}{\bibfnamefont{M.}~\bibnamefont{L{\"u}scher}},
	\bibinfo{journal}{Commun. Math. Phys.} \textbf{\bibinfo{volume}{293}},
	\bibinfo{pages}{899} (\bibinfo{year}{2010}{\natexlab{a}}),
	\eprint{0907.5491}.
	
	\bibitem[{\citenamefont{L{\"u}scher}(2010{\natexlab{b}})}]{Luscher:2010iy}
	\bibinfo{author}{\bibfnamefont{M.}~\bibnamefont{L{\"u}scher}},
	\bibinfo{journal}{JHEP} \textbf{\bibinfo{volume}{08}}, \bibinfo{pages}{071}
	(\bibinfo{year}{2010}{\natexlab{b}}), \bibinfo{note}{[Erratum:
		JHEP03,092(2014)]}, \eprint{1006.4518}.
	
	\bibitem[{\citenamefont{Bonati and D'Elia}(2014)}]{Bonati:2014tqa}
	\bibinfo{author}{\bibfnamefont{C.}~\bibnamefont{Bonati}} \bibnamefont{and}
	\bibinfo{author}{\bibfnamefont{M.}~\bibnamefont{D'Elia}},
	\bibinfo{journal}{Phys. Rev.} \textbf{\bibinfo{volume}{D89}},
	\bibinfo{pages}{105005} (\bibinfo{year}{2014}), \eprint{1401.2441}.
	
	\bibitem[{\citenamefont{Alexandrou et~al.}(2015)\citenamefont{Alexandrou,
			Athenodorou, and Jansen}}]{Alexandrou:2015yba}
	\bibinfo{author}{\bibfnamefont{C.}~\bibnamefont{Alexandrou}},
	\bibinfo{author}{\bibfnamefont{A.}~\bibnamefont{Athenodorou}},
	\bibnamefont{and} \bibinfo{author}{\bibfnamefont{K.}~\bibnamefont{Jansen}},
	\bibinfo{journal}{Phys. Rev.} \textbf{\bibinfo{volume}{D92}},
	\bibinfo{pages}{125014} (\bibinfo{year}{2015}), \eprint{1509.04259}.
	
	\bibitem[{\citenamefont{Panagopoulos and Vicari}(2011)}]{Panagopoulos:2011rb}
	\bibinfo{author}{\bibfnamefont{H.}~\bibnamefont{Panagopoulos}}
	\bibnamefont{and} \bibinfo{author}{\bibfnamefont{E.}~\bibnamefont{Vicari}},
	\bibinfo{journal}{JHEP} \textbf{\bibinfo{volume}{11}}, \bibinfo{pages}{119}
	(\bibinfo{year}{2011}), \eprint{1109.6815}.
	
	\bibitem[{\citenamefont{Bonati et~al.}(2016{\natexlab{b}})\citenamefont{Bonati,
			D'Elia, and Scapellato}}]{Bonati:2015sqt}
	\bibinfo{author}{\bibfnamefont{C.}~\bibnamefont{Bonati}},
	\bibinfo{author}{\bibfnamefont{M.}~\bibnamefont{D'Elia}}, \bibnamefont{and}
	\bibinfo{author}{\bibfnamefont{A.}~\bibnamefont{Scapellato}},
	\bibinfo{journal}{Phys. Rev.} \textbf{\bibinfo{volume}{D93}},
	\bibinfo{pages}{025028} (\bibinfo{year}{2016}{\natexlab{b}}),
	\eprint{1512.01544}.
	
	\bibitem[{\citenamefont{Del~Debbio et~al.}(2004)\citenamefont{Del~Debbio,
			Manca, and Vicari}}]{DelDebbio:2004xh}
	\bibinfo{author}{\bibfnamefont{L.}~\bibnamefont{Del~Debbio}},
	\bibinfo{author}{\bibfnamefont{G.~M.} \bibnamefont{Manca}}, \bibnamefont{and}
	\bibinfo{author}{\bibfnamefont{E.}~\bibnamefont{Vicari}},
	\bibinfo{journal}{Phys. Lett. B} \textbf{\bibinfo{volume}{594}},
	\bibinfo{pages}{315} (\bibinfo{year}{2004}), \eprint{hep-lat/0403001}.
	
	\bibitem[{\citenamefont{Laio et~al.}(2016)\citenamefont{Laio, Martinelli, and
			Sanfilippo}}]{Laio:2015era}
	\bibinfo{author}{\bibfnamefont{A.}~\bibnamefont{Laio}},
	\bibinfo{author}{\bibfnamefont{G.}~\bibnamefont{Martinelli}},
	\bibnamefont{and}
	\bibinfo{author}{\bibfnamefont{F.}~\bibnamefont{Sanfilippo}},
	\bibinfo{journal}{JHEP} \textbf{\bibinfo{volume}{07}}, \bibinfo{pages}{089}
	(\bibinfo{year}{2016}), \eprint{1508.07270}.
	
	\bibitem[{\citenamefont{Flynn et~al.}(2015)\citenamefont{Flynn, Juttner,
			Lawson, and Sanfilippo}}]{Flynn:2015uma}
	\bibinfo{author}{\bibfnamefont{J.}~\bibnamefont{Flynn}},
	\bibinfo{author}{\bibfnamefont{A.}~\bibnamefont{Juttner}},
	\bibinfo{author}{\bibfnamefont{A.}~\bibnamefont{Lawson}}, \bibnamefont{and}
	\bibinfo{author}{\bibfnamefont{F.}~\bibnamefont{Sanfilippo}}
	(\bibinfo{year}{2015}), \eprint{1504.06292}.
	
	\bibitem[{\citenamefont{Hasenbusch}(2017)}]{Hasenbusch:2017unr}
	\bibinfo{author}{\bibfnamefont{M.}~\bibnamefont{Hasenbusch}},
	\bibinfo{journal}{Phys. Rev. D} \textbf{\bibinfo{volume}{96}},
	\bibinfo{pages}{054504} (\bibinfo{year}{2017}), \eprint{1706.04443}.
	
	\bibitem[{\citenamefont{Rossi and Vicari}(1993)}]{Rossi:1993nz}
	\bibinfo{author}{\bibfnamefont{P.}~\bibnamefont{Rossi}} \bibnamefont{and}
	\bibinfo{author}{\bibfnamefont{E.}~\bibnamefont{Vicari}},
	\bibinfo{journal}{Phys. Rev. D} \textbf{\bibinfo{volume}{48}},
	\bibinfo{pages}{3869} (\bibinfo{year}{1993}), \eprint{hep-lat/9301008}.
	
\end{thebibliography}
\end{document}